\documentclass[showkeys,nofootinbib,superscriptaddress,twocolumn]{revtex4}
\usepackage{amsfonts,amssymb,amsthm,bbm,amsmath}
\usepackage{hyperref}

\usepackage{color}

\usepackage{tikz}
\usetikzlibrary{calc}
\usetikzlibrary{decorations.pathmorphing}
\usetikzlibrary{shapes.geometric}
\usetikzlibrary{arrows,decorations.markings}

\newcommand{\C}{{\mathbb C}}
\newcommand{\N}{{\mathbb N}}
\newcommand{\R}{{\mathbb R}}
\newcommand{\Z}{{\mathbb Z}}

\newcommand{\cA}{{\mathcal A}}

\newcommand{\cG}{{\mathcal G}}

\newcommand{\cH}{{\mathcal H}}
\newcommand{\cM}{{\mathcal M}}

\newcommand{\cR}{{\mathcal R}}

\newcommand{\cD}{{\mathcal D}}
\newcommand{\cC}{{\mathcal C}}

\newcommand{\cS}{{\mathcal S}}
\newcommand{\cU}{{\mathcal U}}

\newcommand{\cZ}{{\mathcal Z}}

\newcommand{\SU}{\mathrm{SU}}

\newcommand{\Spin}{\mathrm{Spin}}
\newcommand{\SL}{\mathrm{SL}}
\newcommand{\SO}{\mathrm{SO}}

\newcommand{\be}{\begin{equation}}
\newcommand{\ee}{\end{equation}}
\newcommand{\beq}{\begin{eqnarray}}
\newcommand{\eeq}{\end{eqnarray}}
\newcommand{\bes}{\begin{eqnarray}}
\newcommand{\ees}{\end{eqnarray}}

\newcommand{\su}{{\mathfrak{su}}}

\renewcommand{\sl}{{\mathfrak{sl}}}
\newcommand{\so}{{\mathfrak{so}}}
\newcommand{\osp}{{\mathfrak{osp}}}
\newcommand{\g}{{\mathfrak{g}}}

\newcommand{\la}{\langle}
\newcommand{\ra}{\rangle}

\newcommand{\tr}{{\mathrm{Tr}}}
\newcommand{\f}{\frac}

\def\nn{\nonumber}
\def\pp{\partial}
\newcommand{\w}{\wedge}

\def\rd{\mathrm{d}}

\newcommand{\id}{\mathbb{I}}

\def\vphi{\varphi}
\def\eps{\epsilon}
\def\om{\omega}

\def\vJ{\vec{J}}
\def\vK{\vec{K}}

\colorlet{shadecolor}{gray!40}
\definecolor{blizzardblue}{rgb}{0.67, 0.9, 0.93}

\newcommand{\alink}[4]
{\draw[decoration={markings,mark=at position 0.6 with {\arrow[scale=1.5,>=stealth]{>}}},postaction={decorate}] (#1) -- node[#3,pos=.5]{$#4$}(#2)}
\newcommand{\link}[2]
{\draw[decoration={markings,mark=at position 0.6 with {\arrow[scale=1.5,>=stealth]{>}}},postaction={decorate}] (#1) --(#2)}

\begin{document}

\title{Spinfoam Models for Quantum Gravity: Overview}

\author{{\bf Etera R. Livine}}
\email{etera.livine@ens-lyon.fr}
\affiliation{ENS de Lyon, Laboratoire de Physique, CNRS UMR 5672, Lyon 69007, France}

\date{\today} 
 
\begin{abstract}

In the quest of a physical theory of quantum gravity, spin foam models, or in short spinfoams, propose a well-defined path integral summing over quantized discrete space-time geometries. At the crossroad  of topological quantum field theory, dynamical triangulations, Regge calculus, and loop quantum gravity, this framework provides a non-perturbative and background independent quantization of general relativity. It defines transition amplitudes between quantum states of geometry, and gives a precise picture of the Planck scale geometry with quantized areas and volumes. Gravity in three space-time dimensions is exactly quantized in terms of the Ponzano-Regge state-sum and Turaev-Viro topological invariants. In four space-time dimensions, gravity is formulated as a topological theory, of the BF type, with extra constraints, and hence quantized as a topological state-sum filled with defects. This leads to the Engle-Pereira-Rovelli-Livine (EPRL) spinfoam model, that can be used for explicit quantum gravity computations, for example for resolving the Big Bang singularity by a bounce or in black-to-white hole transition  probability amplitudes.

\end{abstract}


\maketitle  

\makeatletter

\makeatother


\section{Introduction}

Quantum Gravity’s goal is to describe the gravitational interaction at all scales of length and energy, from the microscopic quantum realm to the cosmological dynamics of galaxies. Since the advent of quantum mechanics and general relativity at the beginning of the XXth century, the question of defining a consistent quantum gravity theory remains as one of the biggest challenges of theoretical physics.

On the one hand, general relativity is our best theory to describe classical gravity. It encodes gravity directly in geometry of space-time. Massive objects curve the surrounding space-time, as revealed by the huge body of astrophysical measurements such as the deviation of light, the resulting gravitational lensing effect, the gravitational time dilatation and the cosmological red shift of the expanding universe. And even importantly, it further predicts that the geometry itself can fluctuate, resulting in gravitational waves (as shown by Einstein in 1918), similar to light as electromagnetic waves. This surprising effect, first validated by the decay of binary star systems (historically, the Hulse-Taylor pulsar), has finally been validated directly by the LIGO experiment using laser interferometry, which led to a renewal of observational methods in astrophysics.

On the other hand, quantum mechanics, upgraded to gauge field theory and quantum field theory, explains the physics of atoms, nuclei and particles. It has been thoroughly at both high energy, where the Standard Model of Particle Physics has been validated by the analysis of scattering events in particle accelerators (such as the LEP and the LHC at CERN), and low energy, where it led to the concepts of entanglement and quantum information and to the current boom in quantum computing and quantum technologies. This framework ruling the whole small scale physics predicts that every dynamical degree of physics acquires quantum fluctuations and correlations. Unfortunately, the quantum field theory methods, so successful in describing the electromagnetic and nuclear interactions, fail when applying to general relativity and gravitational waves and lead to apparently uncontrollable quantum fluctuations of the space-time geometry totally in contradiction with our every-day classical universe and known physics. This regime sets a clear limitation and domain of validity of both theories of quantum fields and general relativity. It strongly suggests the necessity of modifications and upgrades of these two frameworks, and pushes us towards a required deepening the concepts of quantum and gravity, at both physical and mathematical levels.

\medskip

Several approaches to the quantum gravity challenge have been developed, such as supergravities, string theory, loop quantum gravity, dynamical triangulations, causal sets, asymptotic safety scenarii, entropic gravity, and non-commutative geometry. Each of them underline specific aspects of the problem and highlight different key physical principles. Among these various approaches, Spinfoams have established themselves as a well-founded legitimate physical theory and an efficient mathematical framework. Set at the crossroads of topological quantum field theory (TQFT), loop quantum gravity, discretized general relativity, causal dynamical triangulations and matrix models for random geometries, spinfoams provide us with a history formulation off the evolution of quantum states of space geometry and define quantum gravity as a path integral over quantized four-dimensional space-time geometries.

Indeed, building on the path integral formulation of quantum mechanics and quantum field theories, one would like to similarly define quantum general relativity as a path integral summing over all possible 4d space-time geometries for a fixed 3d boundary geometry. Technically, this would read
\be
\cA[h]
\equiv
\int_{g_{|\pp \cM} = h} [\cD g]\, e^{-i S_{EH}[g]}
\,,
\ee
where $\cM$ is the four-dimensional space-time manifold, the 3d boundary metric $h$ lives on its three-dimensional boundary $\pp\cM$, and the functional integral is performed over 4d metrics $g$ on $\cM$.
The integrand is given in terms of the Einstein-Hilbert action for general relativity plus suitable boundary terms
(see e.g. \cite{Blau} for a thorough review of general relativity):
\be
l_{P}^{2}S_{EH}[g]
=
\int_{\cM} \sqrt{|\det g|}\,\Big[\cR+\Lambda\Big]
+
\int_{\pp\cM} \sqrt{|\det h|}\,K
\,,
\ee
where $\cR$ is the 4d scalar curvature, defined from the second derivatives of the 4d metric $g$, and $\Lambda$ is the cosmological constant.
The boundary term chosen here is the Hawking-Gibbons-York term, defined in terms of the extrinsic curvature  $K$ of the boundary metric $h$ embedded in the 4d metric $g$.
The pre-factor $l_P=\sqrt{\hbar G /c^{3}}$ is the Planck length defined in terms of the speed of light $c$, Planck constant $\hbar$ and Newton's constant for gravity $G$.
In order to define a transition amplitude between two spatial metrics $h_{i}$ and $h_{f}$, respectively living on an initial hypersurface $\Sigma_{i}$ and a final hypersurface $\Sigma_{f}$, one  applies the formula above to the case $\pp\cM=\Sigma_{i}\cup \Sigma_{f}$, for instance $\cM=\Sigma\times [t_{i},t_{f}]$ in the usual case where the initial and final spatial manifolds are the same $\Sigma_{i}=\Sigma_{f}$.

The standard method, from quantum field theory, to compute such a path integral is to define it perturbatively. One expands the metric $g$  around a suitable physically-relevant background metric, and computes the functional integral perturbatively by integrating over the metric variations  around that background. The path integral is thus expanded in Feynmann diagrams defining the loop corrections to the leading order Gaussian integral. These loop integrals are usually divergent, but one can usually extract meaningful finite parts in a systematic and consistent way. This method is called the renormalization of the path integral. Although this procedure works perfectly for gauge field theories used in particle physics, it unfortunately fails for general relativity and thus can not be used to systematically derive meaningful and reliable quantum corrections to general relativity. This indicates that quantum fluctuations of the geometry at very small scales are much larger than expected and can not be dealt with using the standard toolkit and perturbation methods of quantum field theory. 

One is thus irremediably led to contemplating a non-perturbative definition of the path integral. In order to cure its divergences, one then seeks suitable generalization of its ingredients:
\begin{itemize}
\item a suitable notion of quantized space-time metric $g$ beyond mere perturbations around a background geometry;
\item a suitable notion of quantized boundary conditions, possibly in terms of suitable quantum states of the boundary metric $h$;
\item a suitable probability amplitude extending the exponent of the Einstein-Hilbert action to quantized metrics;
\item a suitable definition of the functional integral over space-time metrics.
\end{itemize}
Spin foams are such a proposal and provide a detailed and precise definition of the local space-time geometry at the Planck scale, with areas, volumes and connections raised to quantum operators (see reviews \cite{Livine:2010zx,Perez:2012wv,Engle:2023qsu}  and introductory textbook \cite{Rovelli:2014ssa}).
It must be noted that, although the development of the framework has historically focussed on general relativity in four space-time dimensions, it also applies to  higher space-times (see e.g.\cite{Freidel:1999rr,Long:2019nkf,Long:2020wuj}) and to classical extension of general relativity such as supergravity theories (see e.g.\cite{Ling:1999gn,Ling:2000dk,Ling:2002ti,Ling:2003yw,Livine:2003hn,Livine:2007dx,Baccetti:2010xd}).

\medskip

We will explain here how spinfoams define space-time histories and transition amplitudes for quantum states of geometry in loop quantum gravity \cite{Rovelli:2014ssa,Reisenberger:1996pu,Perez:2006gja}, how those amplitudes are consistently built from a classical reformulation of general relativity as a constrained topological field theory, how this programme is explicitly realized in three and four space-time dimensions, leading to the Ponzano-Regge model (PR) \cite{Archer:1991rz,Rovelli:1993kc,Freidel:2004vi,Barrett:2008wh} and to the Engle-Pereira-Rovelli-Livine (EPRL) model \cite{Engle:2007wy,Livine:2007ya,Dona:2019dkf}.

We conclude with a summary and short description of active lines of research.


\section{Spin Network Histories}

Spinfoams arised as histories of spin networks in the framework of loop quantum gravity. The name ``spin foam models'' was actually coined by Baez in a 1997 paper \cite{Baez:1997zt} building on the ``sum-over-surfaces'' ideas introduced by Reisenberger and Rovelli \cite{Reisenberger:1996pu}. Loop quantum gravity is a canonical framework for quantum gravity: it defines quantum states of 3d geometry, called {\it spin networks}. Then the evolution in time of those states weaves the fabric of the quantum 4d space-time and defines spinfoams.

\subsection{Spin networks}

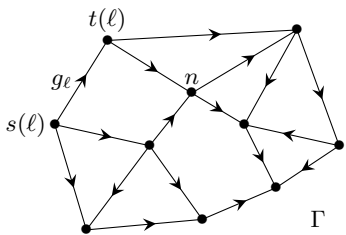
\begin{figure}[h!]

\centering
\begin{tikzpicture}[scale=1.4]

\coordinate(a) at (0,0.2) ;
\coordinate(b) at (.5,1);
\coordinate(c) at (.9,0);
\coordinate(d) at (.3,-.8);
\coordinate(e) at (1.3,.5);
\coordinate(f) at (2.3,1.1);
\coordinate(g) at (1.8,.2);
\coordinate(h) at (2.7,0);
\coordinate(i) at (2.1,-.4);
\coordinate(j) at (1.4,.-.7);

\draw (a) node {$\bullet$} node[left]{$s(\ell)$};
\draw (b) node {$\bullet$} node[above]{$t(\ell)$};
\draw (c) node {$\bullet$} ;
\draw (d) node {$\bullet$};
\draw (e) node {$\bullet$}node[above]{$n$};
\draw (f) node {$\bullet$};
\draw (g) node {$\bullet$};
\draw (h) node {$\bullet$} ++(-0.2,-0.7) node{$\Gamma$};
\draw (i) node {$\bullet$};
\draw (j) node {$\bullet$};

\alink{a}{b}{left}{g_{\ell}};
\link{a}{c};
\link{a}{d};
\link{c}{d};
\link{b}{e};
\link{b}{f};
\link{e}{f};
\link{c}{e};
\link{e}{g};
\link{f}{g};
\link{f}{h};
\link{h}{g};
\link{h}{i};
\link{g}{i};
\link{j}{i};
\link{d}{j};
\link{c}{j};
						
				
\end{tikzpicture}

\caption{Quantum states of geometry are defined as wave-functions of the $\SU(2)$ transport along the links $\ell$ of a graph $\Gamma$, with a gauge invariance under local $\SU(2)$ transformations imposed at every node $n$.}
\label{fig:graph}
\end{figure}

Spin networks are quantum states of 3d geometry. They are defined on graphs, to be embedded in the 3d space, as functions of one $\SU(2)$ group element per graph link. More precisely, considering a finite, oriented, closed graph $\Gamma$, we dress it with group elements $g_{\ell}\in\SU(2)$ on the links $\ell$ and consider functions $\phi$ invariant under $\SU(2)$  transformations at each node $n$ of the graph:
\be
\phi(\{g_{\ell}\}_{\ell\in\Gamma})
=
\phi(\{h_{t(\ell)}g_{\ell}h_{s(\ell)}^{-1}\}_{\ell\in\Gamma})
\,,\,\,
\forall h_{n}\in\SU(2)\,,
\ee
where we write $s(\ell)$ for the source node of the link $\ell$, and $t(\ell)$ for its target node, as illustrated on fig.\ref{fig:graph}.

Intuitively, such functions are interpreted as wave-functions of the transport between space points. Indeed, geometry here is not thought in terms of metric and distances, but in terms of transport\footnotemark{}.
\footnotetext{
More precisely, we use Cartan's formulation of geometry  in terms of vierbein, representing a local basis of vectors, and connection, defining the parallel transport.}
The graph nodes represent space points. The Lie group $\SU(2)$ is the double cover of the orthogonal group $\SO(3)$, so that the group elements $g_{\ell}$ represent the transport between space points, i.e. the change of 3d reference frame along the links between the nodes.  Finally, the invariance under $\SU(2)$  transformations at each node represents the gauge invariance of the states under local change of reference frame at each point of space.

%
The Hilbert space of  quantum states of geometry on that graph is then the $L^{2}$ space of gauge-invariant wave-functions provided with the Haar measure on the Lie group $\SU(2)$:
\be
\cH_{\Gamma}=L^{2}(\SU(2)^{\times \#\ell}/\SU(2)^{\times \# n})\,,
\ee
\be
\la \phi|\widetilde{\phi}\ra_{\Gamma}
=\int_{\SU(2)^{{\times \#\ell}}}\prod_{\ell}\rd g_{\ell}\,
\overline{\phi(\{g_{\ell}\}_{\ell\in\Gamma})}\,\widetilde{\phi}(\{g_{\ell}\}_{\ell\in\Gamma})
\,.
\ee
The full Hilbert space is then defined as the sum of those individual spaces $\cH_{\Gamma}$ over all graphs defined mathematically as a projective limit taking into account  the inclusion of graphs into one another \cite{Ashtekar:1994mh}.

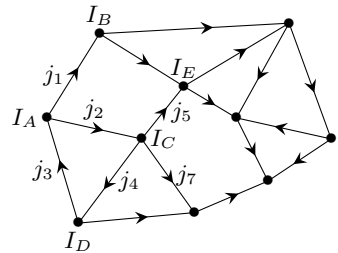
\begin{figure}[h!]

\centering
\begin{tikzpicture}[scale=1.4]

\coordinate(a) at (0,0.2) ;
\coordinate(b) at (.5,1);
\coordinate(c) at (.9,0);
\coordinate(d) at (.3,-.8);
\coordinate(e) at (1.3,.5);
\coordinate(f) at (2.3,1.1);
\coordinate(g) at (1.8,.2);
\coordinate(h) at (2.7,0);
\coordinate(i) at (2.1,-.4);
\coordinate(j) at (1.4,.-.7);

\draw (a) node {$\bullet$} node[left]{$I_{A}$};
\draw (b) node {$\bullet$}node[above]{$I_{B}$};
\draw (c) node {$\bullet$} node[right]{$I_{C}$};
\draw (d) node {$\bullet$} node[below]{$I_{D}$};
\draw (e) node {$\bullet$} node[above]{$I_{E}$};
\draw (f) node {$\bullet$};
\draw (g) node {$\bullet$};
\draw (h) node {$\bullet$};
\draw (i) node {$\bullet$};
\draw (j) node {$\bullet$};

\alink{a}{b}{left}{j_{1}};
\alink{a}{c}{above}{j_{2}};
\alink{d}{a}{left}{j_{3}};
\alink{c}{d}{right}{j_{4}};
\link{b}{e};
\link{b}{f};
\link{e}{f};
\alink{c}{e}{right}{j_{5}};
\link{e}{g};
\link{f}{g};
\link{f}{h};
\link{h}{g};
\link{h}{i};
\link{g}{i};
\link{j}{i};
\link{d}{j};
\alink{c}{j}{right}{j_{7}};

\end{tikzpicture}

\caption{A spin network basis state is labeled by  $\SU(2)$-representations -{\it spins}- $j_{\ell}$  on the graph links and 
$\SU(2)$-invariant tensors -{\it intertwiners}- $I_{n}$ on the graph nodes.
\label{fig:spinnetwork}}
\end{figure}

An orthonormal basis of  $\cH_{\Gamma}$ is defined by using the Plancherel decomposition of functions on $\SU(2)$. This is the equivalent of a Fourier decomposition, where the Fourier modes are given by the Wigner matrices.
This leads the {\it spin network} basis states $|\Gamma,j_{\ell},I_{n}\ra$, labeled by spins $j_{\ell}\in\f\N2$ on the links $\ell$ and intertwiners $I_{n}$ on the nodes $n$, as illustrated on fig.~\ref{fig:spinnetwork}:
\be
\cH_{\Gamma}=\bigoplus_{\{j_{\ell},I_{n}\}_{\ell,n\in\Gamma}}\C\,|\Gamma,j_{\ell},I_{n}\ra\,.
\ee
The spin $j$ labels the irreducible representations of the Lie group $\SU(2)$. The associated  Hilbert space, noted $V^{j}$, has dimension $d_{j}=(2j+1)$. Writing the $\su(2)$ Lie algebra generators as a three-dimensional vector operator $\vJ$, we use the usual basis of $V^{j}$ diagonalizing both the $\su(2)$ Casimir $\vJ^{2}$ and the generator $J_{3}$, thus labeled by the spin $j$ and the magnetic momentum $m$ running by integer step from $-j$ to $+j$:
\be
\vJ^{2}\,|j,m\ra=j(j+1)\,|j,m\ra\,,\quad
J_{3}\,|j,m\ra=m\,|j,m\ra
\,.
\ee
An intertwiner $I_{n}$ at the node $n$ is a $\SU(2)$-invariant map between the tensor product of  incoming spins and the tensor product of  outgoing spins, as illustrated on fig.~\ref{fig:spinnetwork}:
\be
I_{n}:\, \bigotimes_{\ell|t(\ell)=n} V^{j_{\ell}}\,\rightarrow \bigotimes_{\ell|s(\ell)=n} V^{j_{\ell}}
\,,
\ee
thus satisfying $h\circ I_{n}=I_{n}\circ h$ for all $h\in\SU(2)$.
Since spin representations are isomorphic to their complex conjugate, $(V^{j})^{*}\sim V^{j}$,  intertwiners are also understood as singlet states, in the tensor product of all the spins living on the links attached to $n$.
%
By definition of an irreducible representation, bivalent intertwiners only exist if the two spins $j_{1}=j_{2}$ are equal and are then unique. Trivalent intertwiners between three spins exists if and only if those spins satisfy triangular inequalities, $|j_{1}-j_{2}|\le j_{3}\le(j_{1}+j_{2}) $ and are  uniquely given by the Clebsh-Gordan coefficients. From higher node valency, the intertwiner space grows in dimension.
%
%
%

A spin network basis wave-function is then defined by gluing  chosen intertwiners at the nodes together and connecting them by the Wigner matrices of the $\SU(2)$ group elements along the links:
\beq
\vphi^{\{j_{\ell},I_{n}\}}\big{(}\{g_{\ell}\}\big{)}
&=&
\sum_{m_{\ell}^{s,t}}
\prod_{\ell}D^{j_{\ell}}_{m_{\ell}^{t}\,m_{\ell}^{s}}(g_{\ell})
\\
&&
\prod_{n}\la \bigotimes_{\ell|s(\ell)=v} j_{\ell}m_{\ell}^{s} |I_{n}| \bigotimes_{\ell|t(\ell)=n} j_{\ell}m_{\ell}^{t}\ra
\,,\nn
\eeq
where $D^{j}_{mm'}(g)=\la j,m|g|j,m' \ra$ are the matrix elements of  the Wigner matrix $D^{j}(g)$ representing the $\SU(2)$ group element $g$ in the spin-$j$ representation in the magnetic moment basis.


The geometrical interpretation of spin networks in loop quantum gravity comes from the quantization of geometric observables \cite{Rovelli:1995ac}: the metric becomes a differential operator acting on wave-functions.
Intertwiners at the nodes represent excitations of the 3d volume. And spins give the quanta of area of the interfaces between neighbouring chunks of volume \cite{Ashtekar:1996eg,Ashtekar:1997fb}. This is  confirmed by the interpretation of intertwiners as quantized convex polyhedra \cite{Freidel:2009ck,Bianchi:2010gc,Livine:2013tsa} with the spins giving the area of the polyhedra' faces. Spin networks are thus provided with an interpretation as quantized discrete 3d  geometries, actually identified as 3d twisted geometries \cite{Freidel:2010aq} generalizing 3d Regge triangulations \cite{Dittrich:2010ey}.

Explicitly, the spin $j_{\ell}$ gives the area in Planck units of the elementary surface transverse to the link $\ell$:
\be
A[\cS_{\ell}]
=
\gamma\,a_{j_{\ell}}\,l_{P}^{2}
\,.
\ee
The pre-factor $\gamma$ is the Immirzi parameter, which is a new coupling constant (a priori of order 1) controlling the dynamics and fluctuations of geometry in loop quantum gravity. The sequence $a_{j}$ is the area spectrum. It depends on the precise operator ordering used to quantize the area. Originally derived as $a_{j}=\sqrt{j(j+1)}$, another self-consistent choice, also often in spinfoams, is the simpler prescription $a_{j}=j$.
The volume operator is well-defined. Its eigenvalues are understood to scale as $\gamma^{3/2}j^{3/2}l_{P}^{3}$, but it is much harder to diagonalize exactly \cite{Brunnemann:2004xi}. An efficient semi-classical method has recently been developed for computing its eigenvalues \cite{Bianchi:2011ub}.

Spin network basis states are very often referred to merely (but abusively) as ``spin networks''. But one must keep in mind that a spin network is generally an arbitrary superposition of spin network basis states,  thus representing a superposition of quantum geometries.

\medskip

Although spin networks are interpreted in the present context in terms of 3d quantum geometry, following early ideas by Penrose \cite{Penrose1971}, they are actually general tools for gauge field theories, tailor-made for lattice gauge theory. Indeed, the whole procedure developed above, with wave-functions over Lie group elements and basis states labeled by representations and intertwiners, applies to arbitrary semi-simple Lie groups and can even be generalized to quantum groups and spherical categories (e.g. \cite{Barrett:1993zf}).
For instance, a spin network defined on a single loop rooted at a single node, thus a central function $\phi(g)=\phi(hgh^{-1})$, is simply recognized as a Wilson loop observable of gauge field theory, with Fourier modes given by the characters of the irreducible unitary representations of the gauge group.

\subsection{Spinfoams}

\begin{figure}[h!]

\centering

\includegraphics[height=45mm]{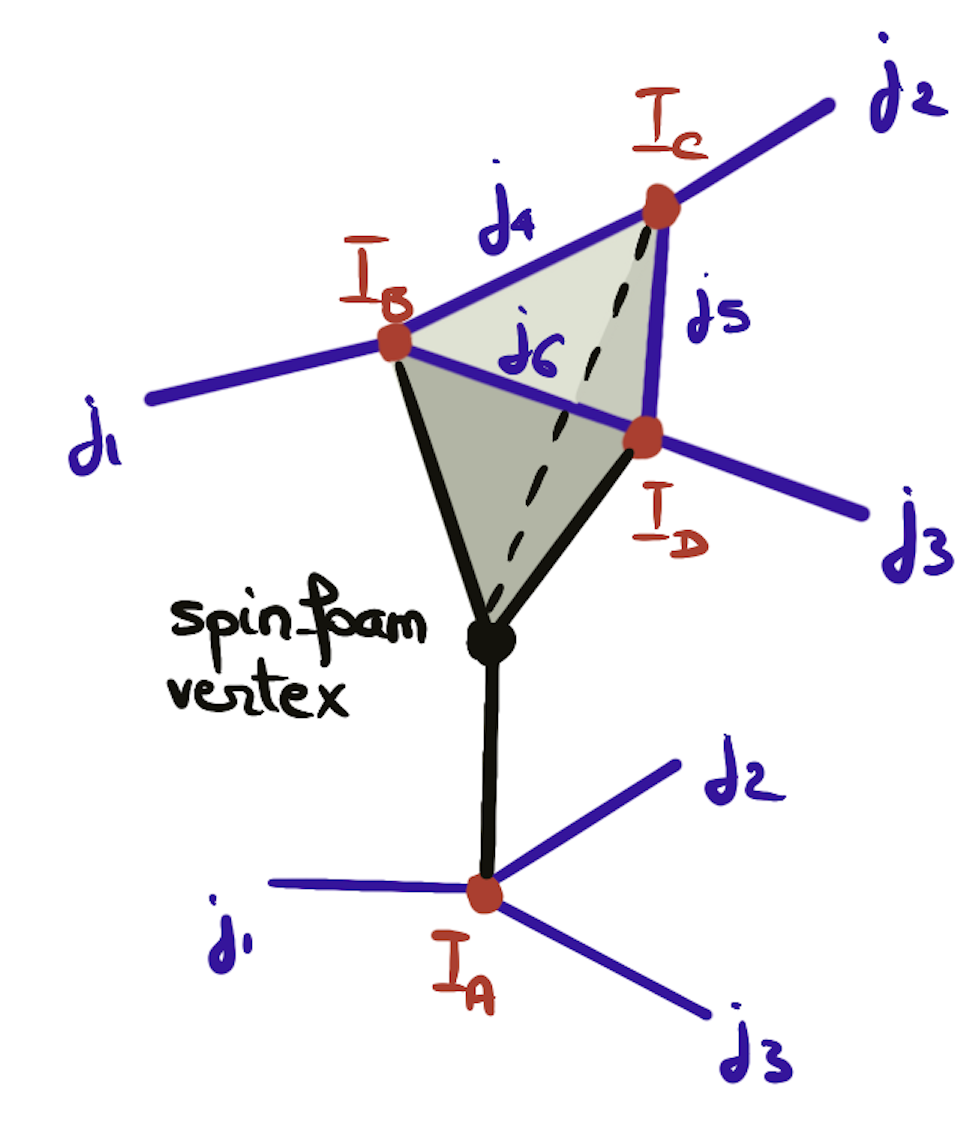} 

\caption{Spinfoam vertex creating a space-time scattering event between quanta of area, carried by spins propagating on 2d worldsheets (faces), and quanta of volumes, carried by intertwiners propagating on worldlines (edges).
\label{fig:spinfoamvertex}}
\end{figure}

Spin networks define quantum states of 3d geometry, and their evolution in time generates the quantum 4d space-time. This is described in terms of {\it spinfoams}.
Graph nodes will evolve in time along a worldline (like a particle) or spinfoam edge, which will be dressed by intertwiners. Graph links will evolve along a 2d worldsheet (like an open string) or spinfoam face, which will carry spins.
Then spinfoam edges and faces can meet at spinfoam vertices, where the graph changes, as illustrated on fig.\ref{fig:spinfoamvertex}. These vertices are actual  events where the spin network states change: they are the space-time points.
We summarize this hierarchy of geometrical cells  in the following table:

\noindent
\begin{center}
\begin{tabular}{|c|c|c|c|}
\hline
spinfoam cell & spatial imprint & algebra & geometry
\\\hline
face & link & spin  & area 
\\\hline
edge & node & intertwiner  & volume 
\\\hline
vertex & transition & amplitude & 4d event
\\\hline
\end{tabular}
\end{center}

\noindent
So spinfoams can be considered as higher dimensional Feynman diagrams, describing the quantum evolution of networks, where  spinfoam vertices play the role of interaction vertices  between the quanta of geometry \cite{Baez:1997zt}. The similarity of the mathematics of spinfoams and Feynam diagrams was explored in \cite{Freidel:2005bb,Baratin:2006yu,Baratin:2006gy}. The actual realization of spinfoams as Feynman diagrams of a quantum field theory is implemented by the {\it group field theory} formalism \cite{Boulatov:1992vp,DePietri:1999bx,Reisenberger:2000zc,Freidel:2005qe,Oriti:2006se}.
%
%

Similarly to Feynman diagrams in quantum field theory, spinfoam amplitudes define probability transition amplitudes for spin network histories. This gives  the local spinfoam model ansatz for a spinfoam cellular complex $\cC$ dressed with representations (or spins) $j_{f}$ on faces and intertwiners $I_{e}$ on edges:
\be
\cA_{\textrm{SF}}[\cC]
=
\prod_{f}\cA_{f}[j_{f}]
\prod_{e}\cA_{e}[j_{f\ni e},I_{e}]
\prod_{v}\cA_{v}[j_{f\ni v},I_{e\ni v}]
\,,
\ee
factorized in terms of face, edge and vertex amplitudes. The face and edge amplitudes can be considered as path integral measure terms, defining the kinematical weights of spins and intertwiners, while the vertex amplitude contains the non-trivial dynamical data encoding the evolution for spin networks.

This ansatz applies to the evolution of spin networks in loop quantum gravity, where one uses representations and intertwiners for $\SU(2)$. But the spinfoam framework is more general and can be applied to arbitrary (semi-simple) Lie groups or finite groups (e.g. \cite{Dittrich:2011zh}). For instance, the Barrett-Crane and EPRL models for 4d quantum gravity is based on the representation theory of the Lorentz group $\SL(2,\C)$, and a non-trivial map (called $Y$-map) between $\SU(2)$ spins and $\SL(2,\C)$ representations needs to be defined to embed loop quantum gravity's spin networks in such spinfoam models \cite{Ding:2010ye,Dupuis:2010jn,Rovelli:2014ssa}. More generally, lattice Yang-Mills gauge theory can be written in the same ansatz \cite{Oeckl:2000hs}.

Then, to compute a spinfoam amplitude for a given boundary state, for instance for transition amplitudes between initial and final spin network states, one fixes (the distribution of) spins and intertwiners  on the boundary and sums over all possible spins and intertwiners in the bulk.

\smallskip

The standard choice of spinfoam amplitudes is given by the BF theory ansatz, i.e. the amplitudes resulting from quantization topological BF field theories \cite{Baez:1997zt}. The face amplitude $\cA_{f}$ is then given by the dimension of representation, $\cA_{f}[j_{f}]=(2j_{f}+1)$ for $\SU(2)$ spins. The edge amplitude $\cA_{e}[I_{e}]$ is the inverse norm of the intertwiner $I_{e}$. The vertex amplitude $\cA_{v}[j_{f},I_{e}]$ is given by the evaluation of the boundary spin networks around the vertex \cite{Perez:2000fs}. More precisely, drawing a 2-sphere around the vertex $v$, the intersection of the spinfoam faces and edges around $v$ define the 1-skeleton of the spinfoam cellular complex around $v$, which we call the {\it boundary graph}. This graph inherits the spin and intertwiner dressing of the spinfoam, defining the boundary spin network $\psi_{\pp v}$. The vertex amplitude is given by the evaluation of this boundary spin network on trivial group elements, which is equivalent to taking the trace over the product of intertwiners. For the $\SU(2)$ Lie group, this gives 3nj-symbols from spin recoupling.
This BF spinfoam ansatz, summarized in the table below, is fully determined by the requirement of topological invariance of the path integral for BF theory, implying that the spinfoam must not depend on the local details of the spinfoam cellular complex but only on its global topology and boundary data \cite{Girelli:2001wr,Bahr:2010bs}. Interestingly, this builds holography directly in the foundations of the spinfoam framework.

\noindent
\begin{center}
\begin{tabular}{|c|c|c|}
\hline
spinfoam cell & algebraic data & BF amplitudes  
\\\hline
face & spin $j_{f}$ & $\cA_{f}=\dim j_{f}$ 
\\\hline
edge &  intertwiner $I_{e}$  & $\cA_{e}=\la I_{e}|I_{e}\ra^{-1}$
\\\hline
vertex & \,boundary spin network\, & $\cA_{v}=\la \id|\psi_{\pp v}\ra$ 
\\\hline
\end{tabular}
\end{center}

\noindent
Most spinfoam models are constructed from the BF ansatz by constraining it, that is restricting the allowed representations and intertwiners, and adding suitable observable insertions, defects or potential terms to the boundary spin network evaluation.

\section{Gravity as a BF Theory}

The basic template for spinfoam path integrals is BF theory, which can be considered as the equivalent for quantum gravity of the harmonic oscillator in quantum mechanics, Yang-Mills in particle physics or Wess-Zumino-Witten gauge theories in conformal field theory.

BF theory is a class of topological gauge theories. On a $D$-dimensional manifold $\cM_{D}$, given a (semi-simple) Lie group $\cG$ with Lie algebra $\g$, the basic fields are a connection 1-form $\om$ valued in $\g$ and a $D-2$ form $B$ also valued in $\g$. The considered action has directly given the name to this class of theories:
\be
S[\om,B]=
\int_{\cM_{D}}
k_{IJ}B^{I}\w F^{J}[\om]
\,,
\ee
for a choice of Killing form on the Lie algebra $\g$ given by the coefficients $k_{IJ}$ in a chosen basis labeled by the indices $I,J$. The 2-form $F[\om]$ is the curvature of the connection:
\be
F^{I}[\om]=\rd \om^{I}+f^{I}_{JK}\om^{J}\w\om^{K}\,,
\ee
in terms of the structure constants $f^{I}_{JK}$ of the Lie algebra.
Its equations of motion are that the curvature and torsion both vanish, respectively $F[\om]=0$ and $\rd_{\om}B=0$. 
In fact, the $B$-field plays the role of a Lagrange multiplier enforcing the flatness of the connection. It is a topological field theory, with no local degrees \cite{Horowitz:1989ng}. Its only physical variables are the global topology of the manifold, and its boundary data, determining the moduli space of flat connections.

Its field theory path integral based on gauge-fixing and BRST techniques leads to Ray-Singer analytic torsion \cite{RaySinger} (see \cite{Bonzom:2010zh,Bonzom:2012mb} for a recent treatment), understood to be equivalent to Reidemeister torsion \cite{Barrett:2008wh}.
The spinfoam framework revisits this quantization and writes the quantum path integral in terms of local amplitudes. 
The idea is straightforward. The path integral imposes the flatness of the connection, handwavingly
\be
Z_{BF}=\int [\cD \om][\cD B]\,e^{i\int B\w F[\om]}
\simeq
\int [\cD \om]\,\delta\big{(}F[\om]\big{)}
\,.
\ee
Now we write a well-defined discrete version of this on the spinfoam 2-complex.
We naturally discretize the connection 1-form $\om$ into group elements $g_{e}$ along the spinfoam edges. We represent the curvature as the holonomy  along the closed loops running around the spinfoam faces and impose that they be trivial. This gives the discrete path integral on a 2-complex $\cC$:
\be
\cZ[\cC]=
\int \prod_{e\in\cC}\rd g_{e}\,
\prod_{f\in\cC}\delta\left(\overleftarrow{\prod_{e\in \pp f}}g_{e}\right)
\,,
\ee
where we use the $\delta$-distribution on the Lie group $\cG$ and we take the product over the oriented group elements around every face. This is the lattice gauge theory formulation of the discrete path integral, e.g. \cite{Reisenberger:1997sk}.
The spinfoam formulation is obtained by decomposing the $\delta$-distribution as the sum over the character of (irreducible unitary) representations of the Lie group. This is the Plancherel formula. For the Lie group $\SU(2)$, it reads:
\be
\delta(g)=\sum_{j\in\f\N2} \dim_{j} \chi_{j}(g)\,,
\ee
where $\chi_{j}(g)=\tr D^{j}(g)$ is the trace of the group element $g$ represented as a matrix in the representation of spin $j$.  Applying this formula to the ordered products of group elements, we obtain group averaging integrals of the type
\be
\int_{\SU(2)} \rd g\,
\prod_{i=1}^{N}D^{j_{i}}(g)
\,,
\ee
which we recognize as the identity on the $\SU(2)$-invariant subspace of the tensor product of spins $\otimes_{i}V^{j_{i}}$. We can thus write them as a decomposition of the identity on a basis of intertwiner states between the $N$ spins:
\be
\int_{\SU(2)} \rd g\,
\prod_{i=1}^{N}D^{j_{i}}(g)
=
\sum_{\lambda}|I^{\lambda}_{\{j_{i}\}}\ra\la I^{\lambda}_{\{j_{i}\}}|
\,,
\ee
where the intertwiner basis is labeled by the index $\lambda$.
So we have one representation $j_{f}$ per spinfoam face, and one intertwiner label $\lambda_{e}$ per spinfoam edge. These intertwiners are contracted around each spinfoam vertex, yielding the expected spinfoam amplitude.
Although we wrote the formulas above for $\cG=\SU(2)$ to alleviate the notations, all the procedure rigorously works for arbitrary semi-simple Lie groups.
In fact, spinfoam models for general relativity in 3+1-dimensional spacetimes (with Lorentzian signature) actually rely on the extension of this lattice gauge theory to the Lorentz group, and require using representations and intertwiners of $\Spin(3,1)\sim\SL(2,\C)$.

We review below how this path integral construction applies to quantum gravity in three and four space-time dimensions.
Indeed,  3d gravity is exactly a BF theory, while 4d gravity (and higher dimensional extensions) can be written as a constrained BF theory, thus quantized as BF theory plus a sea of topological defects. 

%
%


\subsection{3d Quantum Gravity}

3d gravity can be written as a BF gauge theory, with gauge group $\SU(2)$ for a Euclidean space-time signature (+++), and $\SU(1,1)$ for a Lorenztian signature (-++). Here we focus on the Euclidean signature and we will write the resulting spinfoam path integral in terms of $\SU(2)$ representations, but the Lorentzian signature can be treated similarly using the representation theory of $\SU(1,1)$ \cite{Freidel:2000uq,Davids:2000kz,Freidel:2005bb}.

So the first order formulation of 3d gravity, \`a la Cartan, is written in terms of two $\su(2)$-valued 1-forms: the triad  $e=e^{a}J_{a}$ and the connection $\om=\om^{a}J_{a}$. The $J_{a}$'s with $a=1,2,3$ are a basis of  the $\su(2)$ Lie algebra. We choose the standard basis with $[J_{a},J_{b}]=i\eps^{abc} J_{c}$.
The 3d metric is reconstructed as a composite field from the triad as:
\be
g_{\mu\nu}=\eta_{ab}e_{\mu}^{a}e_{\nu}^{b}\,,
\ee
where the internal metric (on the tangent space) $\eta_{ab}=\delta_{ab}$ is the Killing form of the Lie algebra $\su(2)$ in the chosen basis.

The action for Euclidean 3d gravity then reads:
\be
S[e,\om]
=
\int_{\cM_{3d}} e^{a}\w F_{a}[\om] +\f\Lambda{6!} \eps_{abc}e^{a}\w e^{b}\w e^{c}
\,,
\ee
where $F[\om]=\rd\om+\om \w\om$ is the curvature 2-form of the connection.
The first term is the BF Lagrangian, while the second term is a volume term controlled by the cosmological constant $\Lambda$.
This action is invariant under diffeomorphism and $\SU(2)$ gauge transformations acting on the fields as:
\be
e\mapsto geg^{-1}
\,,\quad
\om\mapsto g\om g^{-1}+g\rd g^{-1}
\,.
\ee
As a consequence of those two sets of gauge symmetries, the theory has no physical  local degree of freedom: it is a topological field theory whose partition function depends solely on the global topology of the 3d space-time and on its boundary data.
Since the local fluctuations of the fields are pure gauge, we do not lose information when discretizing the theory. The discrete path integral will thus yield an exact quantization of the theory.

Starting with a non-vanishing cosmological constant $\Lambda=0$, 3d gravity is a pure BF theory and we follow the lattice gauge theory discretization procedure for the path integral as described above. We usually use a 3d triangulation, building the discretized  3d manifold from tetrahedra glued together through shared triangles. As summarized in the table
below, the spinfoam cellular complex $\cC$ is the topological dual of the triangulation $\Delta=\cC^{*}$.
A spinfoam vertex $v$ is dual to tetrahedron $T=v^{*}$ and can be thought as the ``center'' of the tetrahdron.
A spinfoam edge $e$ is dual to a triangle $t=e^{*}$, its two end vertices correspond to the two tetrahedra sharing that triangle.
A spinfoam face $f$ is dual to a edge of the triangulation $\ell=f^{*}$, it has the topology of a disk whose circular boundary goes through all the tetrahedra sharing $\ell$ (cf. fig.\ref{fig:plaquette}).
\noindent
\begin{center}
\begin{tabular}{|c|c|c|}
\hline
spinfoam cell & triangulation & \,algebraic data\,
\\\hline
face $f$ & edge $\ell$ & spin $j_{f}$
\\\hline
edge $e$ &  triangle $t$  & \,intertwiner $I_{e}$\,
\\\hline
vertex $v$ & \,tetrahedron $T$ \,& \,amplitude $\cA_{v}$ \,
\\\hline
\end{tabular}
\end{center}

\begin{figure}[h!]

\centering

\begin{tikzpicture}[scale=1.2]

\coordinate(a) at (0,0) ;
\coordinate(c) at (.3,-2);
\coordinate(b) at (-.3,-1.5);
\coordinate(d) at (-1.3,-1.3);
\coordinate(e) at (1,-.7);
\coordinate(f) at (-0.6,-1.2);
\coordinate(g) at (0.3,-.8);

\draw (f) node {$\bullet$};
\draw (g) node {$\bullet$};

\draw[fill=blizzardblue] (a)--(b)--(c);
\draw[dotted] (a)--(c);
\draw (d)--(a);
\draw (d)--(b);
\draw (d)--(c);
\draw (e)--(a);
\draw (e)--(b);
\draw (e)--(c);

\alink{f}{g}{above left}{g_{t}};

\end{tikzpicture}

\caption{Change of reference frame across the triangular interface between two tetrahedra.
\label{fig:transport-triangle}}
\end{figure}
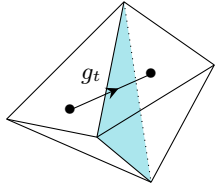

The $\su(2)$ connection is encoded in $\SU(2)$ group elements living on the spinfoam edges $e$, or equivalently attached to the triangles $t=e^{*}$. The geometrical interpretation is as follows: each tetrahedron carries a 3d reference frame and the group element $g_{t}=g_{e^{*}}$ across the triangle $t$ encodes the 3d rotation of reference frames between the two tetrahedra sharing the triangle $t$ (cf fig.\ref{fig:transport-triangle}). 
\begin{figure}[h!]

\centering

\begin{tikzpicture}[scale=1.2]

\coordinate(a) at (0,0) ;
\coordinate(A) at (-0.5,-2);
\coordinate(b1) at (-1,-.4);
\coordinate(b2) at (-.6,-.7);
\coordinate(b3) at (.4,-.9);
\coordinate(b4) at (.8,-.8);

\coordinate(c) at (3,0) ;
\coordinate(c1) at (2.75,-1);
\coordinate(c2) at (2.65,-1.4);
\coordinate(C) at (2.5,-2);
\coordinate(d1) at (2,-.8);
\coordinate(d2) at (2.3,-.6);
\coordinate(d3) at (3.3,-.85);
\coordinate(d4) at (3.4,-1.25);
\coordinate(d5) at (3.1,-1.5);
\coordinate(d6) at (2.2,-1.3);

\draw[thick,red,fill=blizzardblue] (d1)--(d2)--(d3)--node[midway,right]{$e=t^{*}$} (d4)--(d5)--(d6)--(d1);

\draw (d1) node {$\bullet$};
\draw (d1)++(.3,-0.1) node[blue]{$f$};
\draw (d2) node {$\bullet$};
\draw (d3) node {$\bullet$} ;
\draw (d4) node {$\bullet$} ;
\draw (d5) node {$\bullet$}node[below]{$v=T^{*}$};
\draw (d6) node {$\bullet$} ;

\draw[thick] (A)--(a)node[above]{edge $\ell$};
\draw[thick,gray] (a)--(b1)--(A);
\draw[thick,gray] (a)--(b2)--(A);
\draw[thick,gray] (a)--(b3)--(A);
\draw[thick,gray] (a)--(b4)--(A);
\draw[thick,gray] (b1)--(b2)--(b3)--(b4);

\draw (b4)++(.6,0) node{$\rightarrow$};

\draw[thick] (c1)--(c)node[above]{$\ell$};
\draw[thick, dotted] (c1)--(c2);
\draw[thick] (C)--(c2);


\end{tikzpicture}

\caption{A triangulation edge $\ell$ surrounded by the tetrahedra $T$ to which it belongs, and its dual plaquette $f=\ell^{*}$ forming a disk linking the spinfoam vertices $v=T^{*}$ with spinfoam edges $e=t^{*}$.
\label{fig:plaquette}}
\end{figure}
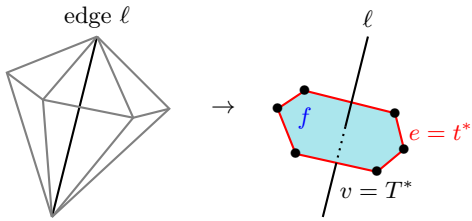

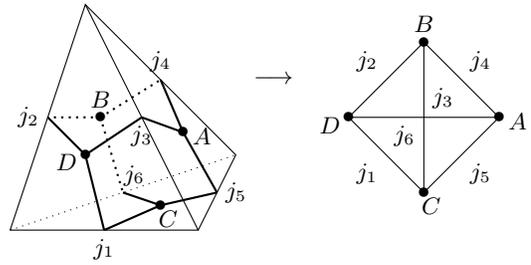
\begin{figure}[h!]

\centering

\begin{tikzpicture}[scale=1]

\coordinate(a) at (0,0) ;
\coordinate(ab) at (1.25,0) ;
\coordinate(b) at (2.5,0);
\coordinate(bc) at (1.75,1.5) ;
\coordinate(ac) at (.5,1.5);
\coordinate(c) at (1,3);
\coordinate(ad) at (1.5,.5);
\coordinate(bd) at (2.75,.5);
\coordinate(cd) at (2,2);
\coordinate(d) at (3,1);

\coordinate(C) at (2,.33);
\draw[thick] (C)--(ab);
\draw[thick] (C)--(bd);
\draw[thick] (C)--(ad);

\coordinate(A) at (2.3,1.3);
\draw[thick] (A)--(bc);
\draw[thick] (A)--(cd);
\draw[thick] (A)--(bd);

\coordinate(B) at (1.2,1.5);
\draw[thick,dotted] (B)--(ac);
\draw[thick,dotted] (B)--(ad);
\draw[thick,dotted] (B)--(cd);

\coordinate(D) at (1,1);
\draw[thick] (D)--(ab);
\draw[thick] (D)--(bc);
\draw[thick] (D)--(ac);

\draw (a)--(b)--(c)--(a);
\draw[dotted] (a)--(d);
\draw (b)--(d)--(c);

\draw (A) node {$\bullet$} ++(.25,-0.05) node{$A$} ;
\draw (B) node {$\bullet$}++(0,.24) node{$B$};
\draw (C) node {$\bullet$}++(.1,-.18) node{$C$} ;
\draw (D) node {$\bullet$}++(-.25,-.1) node{$D$};

\draw(ab)node[below]{$j_{1}$} ;
\draw(ac)node[left]{$j_{2}$} ;
\draw(bc)node[below]{$j_{3}$} ;
\draw(ad)++(0.15,.23) node{$j_{6}$} ;
\draw(bd)node[right]{$j_{5}$} ;
\draw(cd)node[above]{$j_{4}$} ;

\draw (3.5,2) node{$\longrightarrow$};

\coordinate(DD) at (4.5,1.5);
\coordinate(CC) at (5.5,.5);
\coordinate(BB) at (5.5,2.5);
\coordinate(AA) at (6.5,1.5);
\draw (AA) node {$\bullet$} ++(.25,-0.05) node{$A$} ;
\draw (BB) node {$\bullet$}++(0,.24) node{$B$};
\draw (CC) node {$\bullet$}++(.1,-.18) node{$C$} ;
\draw (DD) node {$\bullet$}++(-.25,-.1) node{$D$};
\draw (AA)--node[midway,above right]{$j_{4}$}(BB);
\draw (AA)--node[midway,below right]{$j_{5}$}(CC);
\draw (AA)--node[midway,above right]{$j_{3}$}(DD);
\draw (BB)--node[midway,below left]{$j_{6}$}(CC);
\draw (BB)--node[midway,above left]{$j_{2}$}(DD);
\draw (CC)--node[midway,below left]{$j_{1}$}(DD);

\end{tikzpicture}

\caption{The tetrahedron and its boundary spin network.
\label{fig:6jsymbol}}
\end{figure}

Using the Fourier decomposition of functions over $\SU(2)$ in terms of Wigner matrices, as explained in the previous section, the discretized BF path integral, defined as the integral over the product of $\delta$-distribution enforcing the flatness of the holonomies around every triangulation edge, can be re-written as the product of local amplitudes attached to the tetrahedra \cite{Livine:2010zx,Perez:2012wv}. This is the celebrated Ponzano-Regge formula \cite{osti_4824659}:
\beq
\cZ_{\Delta}&=&
\int \prod_{e\in\Delta^{*}}\rd g_{e}\,
\prod_{f\in\Delta^{*}}\delta\left(\overleftarrow{\prod_{e\in \pp f}}g_{e}\right)
\\
&=&
\sum_{\{j_{\ell}\}} \prod_{\ell}(-1)^{2j_{\ell}} (2j_\ell +1)
\prod_{t}(-1)^{\sum_{\ell\in t} j_{\ell}}
\prod_{T}\{6j\}_{T}
\,,\nn
\eeq
where the Wigner six-j symbol $\{6j\}_{T}$ is the spinfoam amplitude attached to the tetrahedron $T$. It depends on the six spins living on the six edges of the tetrahedron. It is defined as the spin network evaluation on the graph dual to the tetrahedron, where each graph node corresponds to a triangle on the tetrahedron boundary, as drawn on fig.\ref{fig:6jsymbol}.
This is a standard object from spin recoupling theory, constructed as a suitably normalized contraction of the four Clebsh-Gordan coefficients \cite{wiki:6j}. 
We stress that the Ponzano-Regge amplitude is not a thermal partition function but defines a true path integral formulation for 3d quantum gravity.

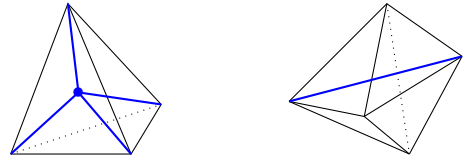
\begin{figure}[h!]

\centering

\begin{tikzpicture}[scale=1]

\coordinate(a) at (0,0) ;
\coordinate(b) at (1.6,0);
\coordinate(c) at (.76,2);
\coordinate(d) at (2,.66);
\coordinate(O) at (.9,.82);

\draw (a)--(b)--(c)--(a);
\draw[dotted] (a)--(d);
\draw (b)--(d)--(c);

\draw (O) node[blue]{$\bullet$};
\draw[blue,thick] (a)--(O)--(b);
\draw[blue,thick] (c)--(O)--(d);

\coordinate(a1) at (5,2) ;
\coordinate(c1) at (5.3,0);
\coordinate(b1) at (4.7,.5);
\coordinate(d1) at (3.7,.7);
\coordinate(e1) at (6,1.3);

\draw (a1)--(b1)--(c1);
\draw[dotted] (a1)--(c1);
\draw (d1)--(a1);
\draw (d1)--(b1);
\draw (d1)--(c1);
\draw (e1)--(a1);
\draw (e1)--(b1);
\draw (e1)--(c1);

\draw[thick, blue] (e1)--(d1);

\end{tikzpicture}
\caption{ 1$\leftrightarrow$4 and 2$\leftrightarrow$3 Pachner moves,  respectively on the left and right, corresponding  to adding or removing a triangulation point or edge (in blue).
\label{fig:Pachner}}
\end{figure}

The spins $j_{\ell}\in\f N2$ define a quantization of edge lengths of the triangulation in Planck units:
\be
L_{\ell}=\left(j_{\ell}+\f12\right)l_{P}
\,.
\ee
This provides the Ponzano-Regge path integral with an interpretation as a quantum geometry: 
states $|j_{\ell},m_{\ell}\ra$ are quantum vectors of length given by their spin $j_{\ell}$, while intertwiners $I_{t}$ represent quantum triangles. Then the probability amplitude of triangles glued together into a tetrahedron is given by the $\{6j\}$-symbol.
This interpretation is confirmed by its large spin behavior:
\be
\left\{
\begin{array}{ccc}
j_{1}&j_{2}&j_{3} \\
j_{4}&j_{5}&j_{6}
\end{array}
\right\}
\underset{j_{k}\gg 1}{\sim}
\f1{\sqrt{12\pi V} } \cos\left(
S_{R}[\{j_{k}\}]
+\f\pi4
\right)\,,
\ee
where $S_{R}=\sum_{k=1}^{6} (j_{k}+\tfrac12)\theta_{k}$ is the Regge action for the tetrahedron. The $\theta_{k}$ are functions of the spins $j_{k}$ and are defined as the external dihedral angles (between the planes of the triangles), so that $S_{R}$ is a discretization of the surface integral of the extrinsic curvature on the tetrahedron boundary.
This asymptotic formula is valid as long as the Cayley determinant, given the squared volume of the tetrahedron in terms of its edge lengths, remains positive and was proven using various methods  \cite{Schulten:1971yv,Roberts:1998zka,Barrett:2002ur,Freidel:2002mj}.
This sets the Ponzano-Regge spinfoam model as a quantized version of Regge calculus (for discretized general relativity in three  dimensions) with quantized lengths.

The key features of this path integral are:
\begin{itemize}

\item it defines a projector on physical states, given by the moduli space of flat connections, as expected for BF theory \cite{Ooguri:1991ni};

\item it is topologically-invariant, in the sense that it des not depend on the details of the bulk triangulation but only on its overall topology; this comes from the Biedenharn-Elliott identity satisfied by the  $\{6j\}$-symbols (also known as the pentagonal  identity in algebra), which implies the invariance of the Ponzano-Regge state-sum under 4$\leftrightarrow$1 and 3$\leftrightarrow$2 Pachner moves,  corresponding respectively to adding or removing a triangulation point or edge, as illustrated on fig.\ref{fig:Pachner};

\item the topological invariance and Biedenharn-Elliott identity translates into recursion relations for the $\{6j\}$-symbols \cite{Schulten:1975yu,Bonzom:2011jh}, which provides a fast iterative algorithm to compute them numerically but is also equivalent to the Wheeler-de Witt equation for 3d quantum gravity (i.e. the quantum version of the Einstein equations) \cite{Bonzom:2011hm};

\item the path integral is actually divergent due to a redundacy of the $\delta$-distributions around each triangulation vertex (due to the Bianchi identity around the dual spinfoam 3d cells); this divergence is due to the symmetry under residual diffeomorphisms acting on the triangulation and can be suitably gauge-fixed, yielding a finite topological invariant \cite{Freidel:2004vi,Bonzom:2010zh,Bonzom:2012mb}; this invariant matches the Reshetikhin-Turaev invariant for the Drinfeld double $\cD\SU(2)$ from the quantization of Chern-Simons theory \cite{Freidel:2004nb}, and the Reidemeister torsion from the BRST quantization of BF theory \cite{Barrett:2008wh};

\item the path integral includes quantum tunnelling through Lorentzian geometries; this exponentially-suppressed geometries come from tetrahedra with negative Cayley determinant, in which case they can not be embedded in a Euclidean signature but are naturally Lorentzian objects \cite{Barrett:1993db}; this allows quantum transition through changes of space-time signature.

\end{itemize}

Having a topological theory provides to a naturally holographic formalism where all the physical information comes either from the bulk topology or from the boundary data. This is made explicit by an exact holographic duality formula with the 2d inhomogeneous Ising model \cite{Bonzom:2015ova}.
The spinfoam amplitude for a triangulation with the 3-ball topology, with a coherent spin network state on its 2-sphere boundary, is the squared inverse of the 2d Ising model living on the boundary graph whose couplings are given by the coherent state parameters.
This formula actually comes a supersymmetry between the fermionic degrees of freedom of the boundary Ising model and the bosonic degrees of freedom of the boundary metric fluctuations. It is an exact formula that holds for arbitrary Ising couplings, even away from criticality and the thermodynamic limit.
This realizes a non-perturbative holographic duality in the deep quantum gravity regime with quantum boundary conditions at finite distance
Other holographic dualities have also been uncovered for various classes of boundary states \cite{Dittrich:2017hnl}, with an apparent link with the Bondi–Metzner–Sachs (BMS) symmetry of gravity for asymptotic flat space-time which still has to be explored further.

\smallskip

A crucial extension of the Ponzano-Regge spinfoam model is the Turaev-Viro model \cite{Turaev:1992hq} implementing a non-vanishing cosmological constant $\Lambda\ne0$ by upgrading the gauge group $\SU(2)$ to the quantum group $\cU_{q}(\su(2))$ with the deformation parameter given by:
\be
q=e^{i\f{2\pi}{k+2}}
\qquad\textrm{with}\quad
k=\f1{G\hbar\sqrt{\Lambda}}
\,.
\ee
We have two regimes for a real cosmological constant: $q$ is real in the Anti-de-Sitter case $\Lambda<0$, and $|q|=1$ for the de Sitter case $\Lambda>0$.
This fits with the quantization of 3d quantum gravity as a Chern-Simons theory as first discussed by Witten \cite{Witten:1988hc,Witten:1988hf,Dittrich:2016typ,Dupuis:2017otn}. The path integral amplitude is formulated in terms of $q$-deformed 6j-symbols, which can be understood as a 3d volume operator acting on the classical 6j-symbols \cite{Livine:2016vhl} and whose large spin asymptotics properly reflect the cosmological constant Lagrangian term.
The model is of particular interest when $k$ is an integer and $q$ is thus a root of unity. In this case, the representation theory of $\cU_{q}(\su(2))$ becomes periodic and is consistently truncated to a finite range of spins $0\le j\le \f k2$. The spinfoam amplitudes are then finite and do not require any gauge-fixing. This yields the Turaev-Viro topological invariants, which is proven to be the squared of the Reshetikhin-Turaev invariant \cite{Reshetikhin:1990pr}. This strongly supports the idea  that the cosmological constant might be quantized in Planck units.

Finally, extension to supergravities for $N=1$ and $N=2$ supersymmetries have also been studied and are formulated in terms of 6j-symbols for the $\Z_{2}$-graded symmetry algebra $\osp(N|2)$ \cite{Livine:2007dx,Baccetti:2010xd}.


\subsection{4d Quantum Gravity}

We proceed with four-dimensional gravity, as for the three-dimensional theory, with the crucial difference that general relativity is not exactly a topological BF theory, but includes a non-trivial potential in the $B$-field leading to physical local fluctuations of the curvature. This is realized by the Plebanski action for 3+1-d general relativity with Lorentzian space-time signature:
\be
S^{4d}[\om,B]
=
\int \eps_{IJKL}B^{IJ}\w F^{KL}[\om] 
+\phi_{IJKL} B^{IJ}\w B^{KL}
\,,
\ee
where $I,J,K,L$ are internal indices on the tangent space. The field $B$ is a $\so(3,1)$-valued 2-form, $\om$ is a $\so(3,1)$-valued connection 1-form and the $\phi_{IJKL}$ are scalar fields with the same symmetry as the 4-form $B^{IJ}\w B^{KL}$:
\be
\phi_{IJKL}=\phi_{KLIJ}=-\phi_{JIKL}=-\phi_{IJLK}
\,.
\ee
This leaves 21 independent scalar field components. We add another condition \cite{Capovilla:2001zi,Livine:2001jt}:
\be
\phi_{IJ}{}^{IJ}
+\f14\left(\gamma-\f1\gamma\right)
\epsilon^{IJKL}\phi_{IJKL}=0
\,,
\ee
which reduces the counting to 20 scalar fields. They play the role of Lagrange multiplier imposing constraints on the $B$-field:
\beq
&&B^{IJ}=E^{IJ}-\f1{2\gamma}\epsilon^{IJKL}E_{KL} 
\\
&&\textrm{with}\quad E^{IJ}\w E^{KL}=\epsilon^{IJKL}\,\f1{4!}\epsilon_{ABCD}E^{AB}\w E^{CD}
\,.\nn
\eeq
These are called {\it simplicity constraints} and require that there exists a tetrad field $e^{I}$ such that \cite{DePietri:1998hnx}:
\be
B^{IJ}=e^{I}\w e^{J} -\f1{2\gamma} \epsilon^{IJKL}e_{K}\w e^{L}
\,.
\ee
A quick counting check gives that the bivector field $B$ has $6\times 6$ independent components, while the tetrad field $e$ has $4\times 4$ independent components. Reducing $B$ to $e$ thus requires 20 Lagrange multipliers, as provided by the $\phi$-fields.
Plugging this expression back in the Plebanski Lagrangian gives back the Palatini-Holst action for general relativity, in its first order formulation \`a la Cartan in terms of tetrad and Lorentz connection \cite{Holst:1995pc}. The coupling constant $\gamma$ is called the Immirzi parameter. It does not enter the Einstein equations for pure gravity, but becomes relevant at the classical level as soon as fermions are coupled to the theory \cite{Perez:2005pm,Freidel:2005sn} and plays a crucial role in the quantum theory.

The Plebanski action has two terms: the  BF Lagrangian plus a potential for the $B$-field. Removing the $B$-potential yields pure BF theory, whose field equations would impose a flat Lorentz connection, thus a vanishing Riemann curvature tensor. The simplicity constraints  remove degrees of freedom of the $B$-fields, thereby relaxing the flatness condition, allowing to get general relativity \cite{Freidel:2012np}. This process can be generalized by playing on the $B$-potential, then yielding a whole class of modified gravity theories with the same number of physical degrees of freedom as general relativity  \cite{Krasnov:2009iy}. Furthermore, it must be noted that the simplicity constraints can be rewritten as a spontaneous symmetry breaking of a $\SO(4,1)$-symmetric Lagrangian in the MacDowell-Mansouri formulation \cite{Freidel:2005ak}.

\smallskip

\begin{figure}[h!]

\centering

\includegraphics[height=18mm]{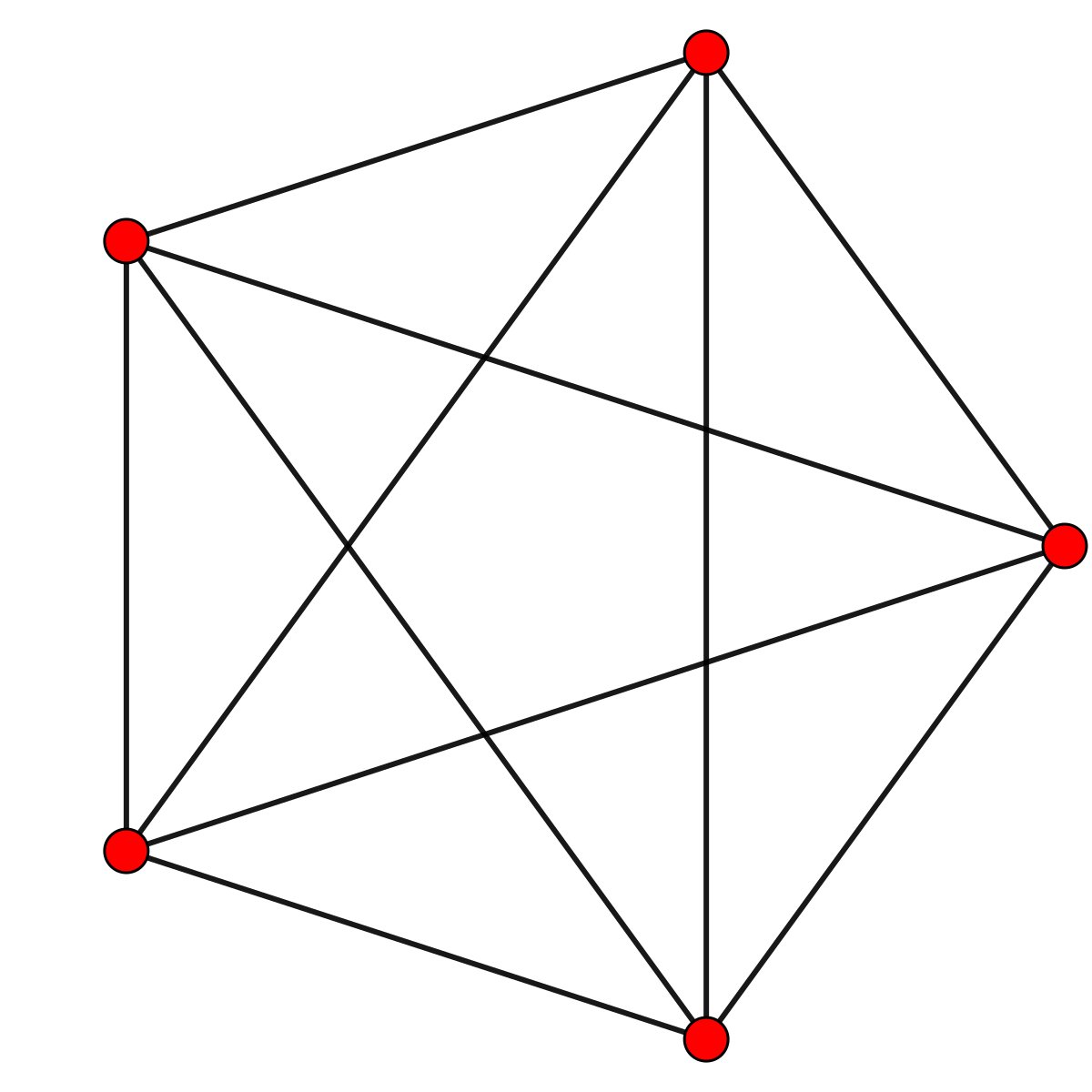}

\caption{The elementary 4d object: a 4-simplex made of 5 points all connected to each other.
\label{fig:4simplex}}
\end{figure}

The quantization strategy is then to first put aside the classical simplicity constraints and define the spinfoam path integral for 4d BF theory, and then impose the simplicity constraints directly at the quantum level.
As for 3d gravity,  the spinfoam cellular complex is chosen as dual to a 4d triangulation. The discrete space-time is thus made of 4-simplices glued together through shared tetrahedra. A 4-simplex is the basic four-dimensional object, defined by 5 non-coplanar points in the 4d manifold, all linked to each other through edges, thus defining 10 triangles and 5 tetrahedra, as illustrated on fig.\ref{fig:4simplex}.

Considering a 4d triangulation $\Delta$, the Lorentz connection $\om$ is discretized into $\Spin(3,1)$ group elements living on  edges in the spinfoam complex $\cC=\Delta^{*}$. These spinfoam edges go through tetrahedra and give the change of Lorentz frames between 4-simplices. The bivector 2-form $B$ is discretized as Lorentz algebra vectors on the triangles, or equivalently or the spinfoam faces. The hierarchy of structures is summarized in the table below:
\noindent
\begin{center}
\begin{tabular}{|c|c|c|c|}
\hline
spinfoam & \,triangulation\, & \,variables\,& \,algebraic data\,
\\\hline
face $f$ & triangle & $B_{f}\in\sl(2,\C)$ & ``spin'' $(n_{f},\rho_{f})$
\\\hline
edge $e$ &  tetrahedron  & $G_{e}\in\SL(2,\C)$ & \,intertwiner $I_{e}$\,
\\\hline
vertex $v$ & \,4-simplex  \,& & \,amplitude $\cA_{v}$ \,
\\\hline
\end{tabular}
\end{center}
This yields the Crane-Yetter topological state-sum (in the case of vanishing quantum deformation) \cite{Crane:1993if,Pfeiffer:2001yd}, with unitary representations of the (double cover of the) Lorentz group $\Spin(3,1)\sim\SL(2,\C)$ dressing the spinfoam faces.
The relevant  representations of the $\SL(2,\C)$ Lie group are from the principal series of unitary irreducible representations, which are labeled by two parameters $(n,\rho)\in\f\N2\times\R$. Choosing the 3d rotation generators $J^{a}$ and the boost generators $K^{a}$  as a basis of the $\sl(2,\C)$ Lie algebra, the Hilbert space of the $\SL(2,\C)$ representations can be decomposed as a direct sum of representations of the $\SU(2)$ subgroup generated by the $J^{a}$'s:
\be
R^{(n,\rho)}=\bigoplus_{j\in n+\N} V^{j}
\,,
\quad
V^{j}=\bigoplus_{-j\le m\le +j}\C|j,m\ra
\,,
\ee
with the Casimir values:
\beq
(\vJ^{2}-\vK^{2})\, &|(n,\rho),j,m\ra&= (n^{2}-\rho^{2}-1)\,|(n,\rho),j,m\ra
\,,\nn\\
\vJ\cdot\vK\, &|(n,\rho),j,m\ra&= n\rho\,|(n,\rho),j,m\ra
\,,\nn\\
\vJ^{2}\, &|(n,\rho),j,m\ra&= j(j+1)\,|(n,\rho),j,m\ra
\,.
\eeq
Details of the action of the $\sl(2,\C)$ generators can be found several textbooks or research papers, e.g. \cite{Alexandrov:2002br}. 

Then we dress the spinfoam edges with  intertwiners between those $\SL(2,\C)$ representations. Finally, the vertex amplitude, attached to a 4-simplex or equivalently to its dual spinfoam vertex, is the boundary spin network evaluation, now given by the contraction of the 5 intertwiners attached to the 5 tetrahedra of the 4-simplex.
This is a topological model, invariant under 4d Pachner moves, adding or removing a vertex, an edge or a face to the triangulation. Its transition amplitude projects onto the moduli space of flat $\SL(2,\C)$ connections.

\smallskip

The goal is then to implement the simplicity constraints directly on the algebraic data dressing the spinfoam cellular complex. These select specific representations and intertwiners of the Lorentz group. The two most developed and most studied models are the Barrett-Crane model (BC) \cite{Barrett:1999qw} and the Engle-Pereira-Rovelli-Livine (EPRL) model \cite{Engle:2007wy}.

The Barrett-Crane (BC) model chooses ``simple'' representations $(n,\rho)=(0,\rho)$ and dressing tetrahedra with the unique Barrett-Crane intertwiner between 4 simple representations such that the tensor product of any pair of representation decomposes also solely into simple representations \cite{Barrett:1999qw}. This is interpreted as quantum tetrahedra with space-like triangles with area $(\rho^{2}+1)\,l_{P}^{2}$. The vertex amplitude for a 4-simplex depends on the representations on its 10 triangles and is defined by the contraction of the 5 Barrett-Crane intertwiners attached to its tetrahedra, yielding a $\{10\rho\}$-symbol \cite{Barrett:1999qw,Perez:2000ec}. Amplitudes can be computed analytically as integrals over $\SL(2,\C)$.

However, despite interesting links with Regge calculus in the large spin asymptotic, the path integral is dominated by degenerate geometries (see \cite{Christensen:2009bi} and references therein). Moreover, one can not make a clear link with the phase space of general relativity or with the spin networks of loop quantum gravity. In particular, there is no Immirzi parameter and the state of a tetrahedron is entirely fixed by its triangle, hence no volume excitation. It nevertheless remains an interesting toy model to test ideas and methods.

The  Engle-Pereira-Rovelli-Livine (EPRL) model was introduced to address all the issues faced by the BC model. Triangles are dressed with a spin $j_{f}$, as expected for spin network histories. The corresponding triangle area $\gamma l_{P}^{2}\sqrt{j_{f}(j_{f}+1)}$ is also consistent with loop quantum gravity. The spin defines a $\SU(2)$ representation, to be embedded in a $\gamma$-simple $\SL(2,\C)$ representation $(n_{f},\rho_{f})=(j_{f},\gamma j_{f})$ \cite{Engle:2007wy}. This embedding map is called the $Y$-map
\cite{Rovelli:2014ssa,Ding:2010ye,Dupuis:2010jn}. One then constructs  tetrahedron states, as intertwiners between $\gamma$-simple representations, leading to the EPRL vertex amplitude defined from $\SL(2,\C)$ group averaging integrals. The reader will find all the mathematical and technical details in \cite{Rovelli:2014ssa,Dona:2019dkf}. The large spin asymptotics relay on coherent state techniques and have been studied in \cite{Barrett:2009mw} and show that the model does indeed realize a path integral over discretized metrics for general relativity.

This is the current spinfoam model for 4d quantum gravity, which is used for quantum cosmology and quantum black hole computations and simulations (see \cite{Christodoulou:2023psv,Han:2024ydv} for recent developments).  Variations of the model exist, e.g.\cite{Banburski:2014cwa}, but are always peaked around the original vertex amplitude.

%
%

\section{Ongoing developments}

To summarize, spinfoams propose a regularized discrete path integral for quantum gravity, well studied in three and four space-time dimensions. Built from topological quantum field theory methods, they provide a description of the geometry at the Planck scale in terms of area, volume and curvature operators. They can be interpreted as a history formalism for loop quantum gravity, providing transition amplitudes between spin network states of quantum geometry. At the end of the day, spinfoams are a well-defined framework for quantum gravity, ripe for analytical and numerical investigation.

Recent important results, beside holographic dualities for 3d quantum gravity, are the spinfoam graviton (i.e. $N$-point correlations between geometric observables over a coherent state peaked on a flat geometry) \cite{Rovelli:2005yj,Livine:2006ab,Alesci:2008ff}, spinfoam cosmology (i.e. the reduction of the spinfoam path integral to homogeneous and isotropic background geometries and their application to cosmology) \cite{Rovelli:2008aa,Livine:2011up,Gozzini:2019nbo}, spinfoam renormalization flow under coarse-graining \cite{Dittrich:2013xwa,Bahr:2016hwc,Steinhaus:2020lgb}.

Current lines of development are three-fold.
First, recent works on mathematical aspects have focussed on  higher gauge symmetries using 2-categories \cite{Asante:2019lki} and the dynamics of defects in spinfoams as extended quantum field theories, which open a very promising interface with the study of topological phases in condensed matter physics.

Then, on formalism development, spinfoams can be reformulated as extended tensor models, or group field theories, allowing for a non-perturbative definition of the sum over space-time triangulations where spinfoam cellular complexes play the role of Feynmann diagrams \cite{Boulatov:1992vp,DePietri:1999bx,Reisenberger:2000zc,Freidel:2005qe,Oriti:2006se}. A side-product of this formalism was the derivation of new results on the large $N$ behavior of random tensor models \cite{Rivasseau:2011hm,Gurau:2012vu}. This leads to an original approach to quantum cosmology, where space-time is modelled as a quantum condensate \cite{Oriti:2016qtz}, and phase diagram for quantum geometries using mean-field methods \cite{Marchetti:2022nrf}.

Finally, to make contact with the semi-classical regime of quantum gravity and its phenomenology, a very promising route is the effective spinfoam approach \cite{Asante:2021zzh,Asante:2022dnj,Dittrich:2022yoo}, investigating the field theory content of the semi-classical limit, coupled to a very active research on the saddle point analysis of the Lorentzian path integral in the space of complexified geometries \cite{Han:2023cen} and the possibility of extensive numerical simulations \cite{Gozzini:2021mex,Dona:2022dxs,Dona:2022yyn}.

\bibliographystyle{bib-style}
\bibliography{biblioQG}

\providecommand{\href}[2]{#2}\begingroup\raggedright\begin{thebibliography}{100}

\bibitem{Blau}
M.~Blau, ``Lecture Notes on General Relativity,'' 2023.
\newblock Universit\"at Bern, available on
  \url{http://www.blau.itp.unibe.ch/GRLecturenotes.html}.

\bibitem{Livine:2010zx}
E.~R. Livine, ``{The Spinfoam Framework for Quantum Gravity},''
  \href{http://arXiv.org/abs/1101.5061}{{\texttt{arXiv:1101.5061}}}.
  Habilitation Thesis 2010, ENS de Lyon (France).

\bibitem{Perez:2012wv}
A.~Perez, ``{The Spin Foam Approach to Quantum Gravity},'' Living Rev. Rel.
  {\bf 16} (2013) 3,
  \href{http://arXiv.org/abs/1205.2019}{{\texttt{arXiv:1205.2019}}}.

\bibitem{Engle:2023qsu}
J.~Engle and S.~Speziale, {\em {Spin Foams: Foundations}}.
\newblock 2023.
\newblock \href{http://arXiv.org/abs/2310.20147}{{\texttt{arXiv:2310.20147}}}.

\bibitem{Rovelli:2014ssa}
C.~Rovelli and F.~Vidotto, {\em {Covariant Loop Quantum Gravity}: {An
  Elementary Introduction to Quantum Gravity and Spinfoam Theory}}.
\newblock Cambridge Monographs on Mathematical Physics. Cambridge University
  Press, 11, 2014.

\bibitem{Freidel:1999rr}
L.~Freidel, K.~Krasnov, and R.~Puzio, ``{BF description of higher dimensional
  gravity theories},'' Adv. Theor. Math. Phys. {\bf 3} (1999) 1289--1324,
  \href{http://arXiv.org/abs/hep-th/9901069}{{\texttt{arXiv:hep-th/9901069}}}.

\bibitem{Long:2019nkf}
G.~Long, C.-Y. Lin, and Y.~Ma, ``{Coherent intertwiner solution of simplicity
  constraint in all dimensional loop quantum gravity},'' Phys. Rev. D {\bf 100}
  (2019), no.~6, 064065,
  \href{http://arXiv.org/abs/1906.06534}{{\texttt{arXiv:1906.06534}}}.

\bibitem{Long:2020wuj}
G.~Long and Y.~Ma, ``{General geometric operators in all dimensional loop
  quantum gravity},'' Phys. Rev. D {\bf 101} (2020), no.~8, 084032,
  \href{http://arXiv.org/abs/2003.03952}{{\texttt{arXiv:2003.03952}}}.

\bibitem{Ling:1999gn}
Y.~Ling and L.~Smolin, ``{Supersymmetric spin networks and quantum
  supergravity},'' Phys. Rev. D {\bf 61} (2000) 044008,
  \href{http://arXiv.org/abs/hep-th/9904016}{{\texttt{arXiv:hep-th/9904016}}}.

\bibitem{Ling:2000dk}
Y.~Ling and L.~Smolin, ``{Eleven-dimensional supergravity as a constrained
  topological field theory},'' Nucl. Phys. B {\bf 601} (2001) 191--208,
  \href{http://arXiv.org/abs/hep-th/0003285}{{\texttt{arXiv:hep-th/0003285}}}.

\bibitem{Ling:2002ti}
Y.~Ling, {\em {Extending loop quantum gravity to supergravity}}.
\newblock {PhD thesis}, {Imperial College (London, UK)}, 2002.

\bibitem{Ling:2003yw}
Y.~Ling, R.-S. Tung, and H.-Y. Guo, ``{Supergravity and Yang-Mills theories as
  generalized topological fields with constraints},'' Phys. Rev. D {\bf 70}
  (2004) 044045,
  \href{http://arXiv.org/abs/hep-th/0310141}{{\texttt{arXiv:hep-th/0310141}}}.

\bibitem{Livine:2003hn}
E.~R. Livine and R.~Oeckl, ``{Three-dimensional Quantum Supergravityand
  Supersymmetric Spin Foam Models},'' Adv. Theor. Math. Phys. {\bf 7} (2003),
  no.~6, 951--1001,
  \href{http://arXiv.org/abs/hep-th/0307251}{{\texttt{arXiv:hep-th/0307251}}}.

\bibitem{Livine:2007dx}
E.~R. Livine and J.~P. Ryan, ``{N=2 supersymmetric spin foams in three
  dimensions},'' Class. Quant. Grav. {\bf 25} (2008) 175014,
  \href{http://arXiv.org/abs/0710.3540}{{\texttt{arXiv:0710.3540}}}.

\bibitem{Baccetti:2010xd}
V.~Baccetti, E.~R. Livine, and J.~P. Ryan, ``{The Particle interpretation of N
  = 1 supersymmetric spin foams},'' Class. Quant. Grav. {\bf 27} (2010) 225022,
  \href{http://arXiv.org/abs/1004.0672}{{\texttt{arXiv:1004.0672}}}.

\bibitem{Reisenberger:1996pu}
M.~P. Reisenberger and C.~Rovelli, ``{'Sum over surfaces' form of loop quantum
  gravity},'' Phys. Rev. D {\bf 56} (1997) 3490--3508,
  \href{http://arXiv.org/abs/gr-qc/9612035}{{\texttt{arXiv:gr-qc/9612035}}}.

\bibitem{Perez:2006gja}
A.~Perez, ``{The Spin-foam-representation of LQG},''
  \href{http://arXiv.org/abs/gr-qc/0601095}{{\texttt{arXiv:gr-qc/0601095}}}.

\bibitem{Archer:1991rz}
F.~Archer and R.~M. Williams, ``{The Turaev-Viro state sum model and
  three-dimensional quantum gravity},'' Phys. Lett. B {\bf 273} (1991)
  438--444.

\bibitem{Rovelli:1993kc}
C.~Rovelli, ``{The Basis of the Ponzano-Regge-Turaev-Viro-Ooguri quantum
  gravity model in the loop representation basis},'' Phys. Rev. D {\bf 48}
  (1993) 2702--2707,
  \href{http://arXiv.org/abs/hep-th/9304164}{{\texttt{arXiv:hep-th/9304164}}}.

\bibitem{Freidel:2004vi}
L.~Freidel and D.~Louapre, ``{Ponzano-Regge model revisited I: Gauge fixing,
  observables and interacting spinning particles},'' Class. Quant. Grav. {\bf
  21} (2004) 5685--5726,
  \href{http://arXiv.org/abs/hep-th/0401076}{{\texttt{arXiv:hep-th/0401076}}}.

\bibitem{Barrett:2008wh}
J.~W. Barrett and I.~Naish-Guzman, ``{The Ponzano-Regge model},'' Class. Quant.
  Grav. {\bf 26} (2009) 155014,
  \href{http://arXiv.org/abs/0803.3319}{{\texttt{arXiv:0803.3319}}}.

\bibitem{Engle:2007wy}
J.~Engle, E.~Livine, R.~Pereira, and C.~Rovelli, ``{LQG vertex with finite
  Immirzi parameter},'' Nucl. Phys. B {\bf 799} (2008) 136--149,
  \href{http://arXiv.org/abs/0711.0146}{{\texttt{arXiv:0711.0146}}}.

\bibitem{Livine:2007ya}
E.~R. Livine and S.~Speziale, ``{Consistently Solving the Simplicity
  Constraints for Spinfoam Quantum Gravity},'' EPL {\bf 81} (2008), no.~5,
  50004, \href{http://arXiv.org/abs/0708.1915}{{\texttt{arXiv:0708.1915}}}.

\bibitem{Dona:2019dkf}
P.~Don\`a, M.~Fanizza, G.~Sarno, and S.~Speziale, ``{Numerical study of the
  Lorentzian Engle-Pereira-Rovelli-Livine spin foam amplitude},'' Phys. Rev. D
  {\bf 100} (2019), no.~10, 106003,
  \href{http://arXiv.org/abs/1903.12624}{{\texttt{arXiv:1903.12624}}}.

\bibitem{Baez:1997zt}
J.~C. Baez, ``{Spin foam models},'' Class. Quant. Grav. {\bf 15} (1998)
  1827--1858,
  \href{http://arXiv.org/abs/gr-qc/9709052}{{\texttt{arXiv:gr-qc/9709052}}}.

\bibitem{Ashtekar:1994mh}
A.~Ashtekar and J.~Lewandowski, ``{Projective techniques and functional
  integration for gauge theories},'' J. Math. Phys. {\bf 36} (1995) 2170--2191,
  \href{http://arXiv.org/abs/gr-qc/9411046}{{\texttt{arXiv:gr-qc/9411046}}}.

\bibitem{Rovelli:1995ac}
C.~Rovelli and L.~Smolin, ``{Spin networks and quantum gravity},'' Phys. Rev. D
  {\bf 52} (1995) 5743--5759,
  \href{http://arXiv.org/abs/gr-qc/9505006}{{\texttt{arXiv:gr-qc/9505006}}}.

\bibitem{Ashtekar:1996eg}
A.~Ashtekar and J.~Lewandowski, ``{Quantum theory of geometry. 1: Area
  operators},'' Class. Quant. Grav. {\bf 14} (1997) A55--A82,
  \href{http://arXiv.org/abs/gr-qc/9602046}{{\texttt{arXiv:gr-qc/9602046}}}.

\bibitem{Ashtekar:1997fb}
A.~Ashtekar and J.~Lewandowski, ``{Quantum theory of geometry. 2. Volume
  operators},'' Adv. Theor. Math. Phys. {\bf 1} (1998) 388--429,
  \href{http://arXiv.org/abs/gr-qc/9711031}{{\texttt{arXiv:gr-qc/9711031}}}.

\bibitem{Freidel:2009ck}
L.~Freidel and E.~R. Livine, ``{The Fine Structure of SU(2) Intertwiners from
  U(N) Representations},'' J. Math. Phys. {\bf 51} (2010) 082502,
  \href{http://arXiv.org/abs/0911.3553}{{\texttt{arXiv:0911.3553}}}.

\bibitem{Bianchi:2010gc}
E.~Bianchi, P.~Dona, and S.~Speziale, ``{Polyhedra in loop quantum gravity},''
  Phys. Rev. D {\bf 83} (2011) 044035,
  \href{http://arXiv.org/abs/1009.3402}{{\texttt{arXiv:1009.3402}}}.

\bibitem{Livine:2013tsa}
E.~R. Livine, ``{Deformations of Polyhedra and Polygons by the Unitary
  Group},'' J. Math. Phys. {\bf 54} (2013) 123504,
  \href{http://arXiv.org/abs/1307.2719}{{\texttt{arXiv:1307.2719}}}.

\bibitem{Freidel:2010aq}
L.~Freidel and S.~Speziale, ``{Twisted geometries: A geometric parametrisation
  of SU(2) phase space},'' Phys. Rev. D {\bf 82} (2010) 084040,
  \href{http://arXiv.org/abs/1001.2748}{{\texttt{arXiv:1001.2748}}}.

\bibitem{Dittrich:2010ey}
B.~Dittrich and J.~P. Ryan, ``{Simplicity in simplicial phase space},'' Phys.
  Rev. D {\bf 82} (2010) 064026,
  \href{http://arXiv.org/abs/1006.4295}{{\texttt{arXiv:1006.4295}}}.

\bibitem{Brunnemann:2004xi}
J.~Brunnemann and T.~Thiemann, ``{Simplification of the spectral analysis of
  the volume operator in loop quantum gravity},'' Class. Quant. Grav. {\bf 23}
  (2006) 1289--1346,
  \href{http://arXiv.org/abs/gr-qc/0405060}{{\texttt{arXiv:gr-qc/0405060}}}.

\bibitem{Bianchi:2011ub}
E.~Bianchi and H.~M. Haggard, ``{Discreteness of the volume of space from
  Bohr-Sommerfeld quantization},'' Phys. Rev. Lett. {\bf 107} (2011) 011301,
  \href{http://arXiv.org/abs/1102.5439}{{\texttt{arXiv:1102.5439}}}.

\bibitem{Penrose1971}
R.~Penrose, ``{Angular momentum: an approach to combinatorial space-time},''.
  available online at \url{https://math.ucr.edu/home/baez/penrose/}.

\bibitem{Barrett:1993zf}
J.~W. Barrett and B.~W. Westbury, ``{Spherical categories},'' Adv. Math. {\bf
  143} (1999) 357--375,
  \href{http://arXiv.org/abs/hep-th/9310164}{{\texttt{arXiv:hep-th/9310164}}}.

\bibitem{Freidel:2005bb}
L.~Freidel and E.~R. Livine, ``{Ponzano-Regge model revisited III: Feynman
  diagrams and effective field theory},'' Class. Quant. Grav. {\bf 23} (2006)
  2021--2062,
  \href{http://arXiv.org/abs/hep-th/0502106}{{\texttt{arXiv:hep-th/0502106}}}.

\bibitem{Baratin:2006yu}
A.~Baratin and L.~Freidel, ``{Hidden Quantum Gravity in 3-D Feynman
  diagrams},'' Class. Quant. Grav. {\bf 24} (2007) 1993--2026,
  \href{http://arXiv.org/abs/gr-qc/0604016}{{\texttt{arXiv:gr-qc/0604016}}}.

\bibitem{Baratin:2006gy}
A.~Baratin and L.~Freidel, ``{Hidden Quantum Gravity in 4-D Feynman diagrams:
  Emergence of spin foams},'' Class. Quant. Grav. {\bf 24} (2007) 2027--2060,
  \href{http://arXiv.org/abs/hep-th/0611042}{{\texttt{arXiv:hep-th/0611042}}}.

\bibitem{Boulatov:1992vp}
D.~V. Boulatov, ``{A Model of three-dimensional lattice gravity},'' Mod. Phys.
  Lett. A {\bf 7} (1992) 1629--1646,
  \href{http://arXiv.org/abs/hep-th/9202074}{{\texttt{arXiv:hep-th/9202074}}}.

\bibitem{DePietri:1999bx}
R.~De~Pietri, L.~Freidel, K.~Krasnov, and C.~Rovelli, ``{Barrett-Crane model
  from a Boulatov-Ooguri field theory over a homogeneous space},'' Nucl. Phys.
  B {\bf 574} (2000) 785--806,
  \href{http://arXiv.org/abs/hep-th/9907154}{{\texttt{arXiv:hep-th/9907154}}}.

\bibitem{Reisenberger:2000zc}
M.~P. Reisenberger and C.~Rovelli, ``{Space-time as a Feynman diagram: The
  Connection formulation},'' Class. Quant. Grav. {\bf 18} (2001) 121--140,
  \href{http://arXiv.org/abs/gr-qc/0002095}{{\texttt{arXiv:gr-qc/0002095}}}.

\bibitem{Freidel:2005qe}
L.~Freidel, ``{Group field theory: An Overview},'' Int. J. Theor. Phys. {\bf
  44} (2005) 1769--1783,
  \href{http://arXiv.org/abs/hep-th/0505016}{{\texttt{arXiv:hep-th/0505016}}}.

\bibitem{Oriti:2006se}
D.~Oriti, ``{The Group field theory approach to quantum gravity},''
  \href{http://arXiv.org/abs/gr-qc/0607032}{{\texttt{arXiv:gr-qc/0607032}}}.

\bibitem{Dittrich:2011zh}
B.~Dittrich, F.~C. Eckert, and M.~Martin-Benito, ``{Coarse graining methods for
  spin net and spin foam models},'' New J. Phys. {\bf 14} (2012) 035008,
  \href{http://arXiv.org/abs/1109.4927}{{\texttt{arXiv:1109.4927}}}.

\bibitem{Ding:2010ye}
Y.~Ding and C.~Rovelli, ``{Physical boundary Hilbert space and volume operator
  in the Lorentzian new spin-foam theory},'' Class. Quant. Grav. {\bf 27}
  (2010) 205003,
  \href{http://arXiv.org/abs/1006.1294}{{\texttt{arXiv:1006.1294}}}.

\bibitem{Dupuis:2010jn}
M.~Dupuis and E.~R. Livine, ``{Lifting SU(2) Spin Networks to Projected Spin
  Networks},'' Phys. Rev. D {\bf 82} (2010) 064044,
  \href{http://arXiv.org/abs/1008.4093}{{\texttt{arXiv:1008.4093}}}.

\bibitem{Oeckl:2000hs}
R.~Oeckl and H.~Pfeiffer, ``{The Dual of pure nonAbelian lattice gauge theory
  as a spin foam model},'' Nucl. Phys. B {\bf 598} (2001) 400--426,
  \href{http://arXiv.org/abs/hep-th/0008095}{{\texttt{arXiv:hep-th/0008095}}}.

\bibitem{Perez:2000fs}
A.~Perez and C.~Rovelli, ``{A Spin foam model without bubble divergences},''
  Nucl. Phys. B {\bf 599} (2001) 255--282,
  \href{http://arXiv.org/abs/gr-qc/0006107}{{\texttt{arXiv:gr-qc/0006107}}}.

\bibitem{Girelli:2001wr}
F.~Girelli, R.~Oeckl, and A.~Perez, ``{Spin foam diagrammatics and topological
  invariance},'' Class. Quant. Grav. {\bf 19} (2002) 1093--1108,
  \href{http://arXiv.org/abs/gr-qc/0111022}{{\texttt{arXiv:gr-qc/0111022}}}.

\bibitem{Bahr:2010bs}
B.~Bahr, F.~Hellmann, W.~Kaminski, M.~Kisielowski, and J.~Lewandowski,
  ``{Operator Spin Foam Models},'' Class. Quant. Grav. {\bf 28} (2011) 105003,
  \href{http://arXiv.org/abs/1010.4787}{{\texttt{arXiv:1010.4787}}}.

\bibitem{Horowitz:1989ng}
G.~T. Horowitz, ``{Exactly Soluble Diffeomorphism Invariant Theories},''
  Commun. Math. Phys. {\bf 125} (1989) 417.

\bibitem{RaySinger}
D.~B. Ray and I.~M. Singer, ``Analytic Torsion for Complex Manifolds,'' Annals
  of Mathematics {\bf 98} (1973), no.~1, 154--177.

\bibitem{Bonzom:2010zh}
V.~Bonzom and M.~Smerlak, ``{Bubble divergences from twisted cohomology},''
  Commun. Math. Phys. {\bf 312} (2012) 399--426,
  \href{http://arXiv.org/abs/1008.1476}{{\texttt{arXiv:1008.1476}}}.

\bibitem{Bonzom:2012mb}
V.~Bonzom and M.~Smerlak, ``{Gauge symmetries in spinfoam gravity: the case for
  'cellular quantization'},'' Phys. Rev. Lett. {\bf 108} (2012) 241303,
  \href{http://arXiv.org/abs/1201.4996}{{\texttt{arXiv:1201.4996}}}.

\bibitem{Reisenberger:1997sk}
M.~P. Reisenberger, ``{A Lattice world sheet sum for 4-d Euclidean general
  relativity},''
  \href{http://arXiv.org/abs/gr-qc/9711052}{{\texttt{arXiv:gr-qc/9711052}}}.

\bibitem{Freidel:2000uq}
L.~Freidel, ``{A Ponzano-Regge model of Lorentzian 3-dimensional gravity},''
  Nucl. Phys. B Proc. Suppl. {\bf 88} (2000) 237--240,
  \href{http://arXiv.org/abs/gr-qc/0102098}{{\texttt{arXiv:gr-qc/0102098}}}.

\bibitem{Davids:2000kz}
S.~Davids, {\em {A State sum model for (2+1) Lorentzian quantum gravity}}.
\newblock PhD thesis, Nottingham University (UK), 2000.
\newblock
  \href{http://arXiv.org/abs/gr-qc/0110114}{{\texttt{arXiv:gr-qc/0110114}}}.

\bibitem{osti_4824659}
G.~Ponzano and T.~Regge, ``Semiclassical Limit of Racah Coefficients,'' pp 1-58
  of Spectroscopic and Group Theoretical Methods in Physics. Block, F. (ed.).
  New York, John Wiley and Sons, Inc., 1968. (10, 1969).

\bibitem{wiki:6j}
Wikipedia, ``{6-j symbol} --- {W}ikipedia{,} The Free Encyclopedia.''
  \url{https://en.wikipedia.org/wiki/6-j_symbol}, 2023.
\newblock [Online; accessed 24-November-2023].

\bibitem{Schulten:1971yv}
K.~Schulten and R.~G. Gordon, ``{Semiclassical Approximations to 3j and 6j
  coefficients for Quantum Mechanical Coupling of Angular Momenta},'' J. Math.
  Phys. {\bf 16} (1975) 1971--1988.

\bibitem{Roberts:1998zka}
J.~Roberts, ``{Classical 6j-symbols and the tetrahedron},'' Geom. Topol. {\bf
  3} (1999), no.~1, 21--66,
  \href{http://arXiv.org/abs/math-ph/9812013}{{\texttt{arXiv:math-ph/9812013}}}.

\bibitem{Barrett:2002ur}
J.~W. Barrett and C.~M. Steele, ``{Asymptotics of relativistic spin
  networks},'' Class. Quant. Grav. {\bf 20} (2003) 1341--1362,
  \href{http://arXiv.org/abs/gr-qc/0209023}{{\texttt{arXiv:gr-qc/0209023}}}.

\bibitem{Freidel:2002mj}
L.~Freidel and D.~Louapre, ``{Asymptotics of 6j and 10j symbols},'' Class.
  Quant. Grav. {\bf 20} (2003) 1267--1294,
  \href{http://arXiv.org/abs/hep-th/0209134}{{\texttt{arXiv:hep-th/0209134}}}.

\bibitem{Ooguri:1991ni}
H.~Ooguri, ``{Partition functions and topology changing amplitudes in the 3-D
  lattice gravity of Ponzano and Regge},'' Nucl. Phys. B {\bf 382} (1992)
  276--304,
  \href{http://arXiv.org/abs/hep-th/9112072}{{\texttt{arXiv:hep-th/9112072}}}.

\bibitem{Schulten:1975yu}
K.~Schulten and R.~G. Gordon, ``{Exact Recursive Evaluation of 3J and 6J
  Coefficients for Quantum Mechanical Coupling of Angular Momenta},'' J. Math.
  Phys. {\bf 16} (1975) 1961--1970.

\bibitem{Bonzom:2011jh}
V.~Bonzom and E.~R. Livine, ``{A New Recursion Relation for the 6j-Symbol},''
  Annales Henri Poincare {\bf 13} (2012) 1083--1099,
  \href{http://arXiv.org/abs/1103.3415}{{\texttt{arXiv:1103.3415}}}.

\bibitem{Bonzom:2011hm}
V.~Bonzom and L.~Freidel, ``{The Hamiltonian constraint in 3d Riemannian loop
  quantum gravity},'' Class. Quant. Grav. {\bf 28} (2011) 195006,
  \href{http://arXiv.org/abs/1101.3524}{{\texttt{arXiv:1101.3524}}}.

\bibitem{Freidel:2004nb}
L.~Freidel and D.~Louapre, ``{Ponzano-Regge model revisited II: Equivalence
  with Chern-Simons},''
  \href{http://arXiv.org/abs/gr-qc/0410141}{{\texttt{arXiv:gr-qc/0410141}}}.

\bibitem{Barrett:1993db}
J.~W. Barrett and T.~J. Foxon, ``{Semiclassical limits of simplicial quantum
  gravity},'' Class. Quant. Grav. {\bf 11} (1994) 543--556,
  \href{http://arXiv.org/abs/gr-qc/9310016}{{\texttt{arXiv:gr-qc/9310016}}}.

\bibitem{Bonzom:2015ova}
V.~Bonzom, F.~Costantino, and E.~R. Livine, ``{Duality between Spin networks
  and the 2D Ising model},'' Commun. Math. Phys. {\bf 344} (2016), no.~2,
  531--579,
  \href{http://arXiv.org/abs/1504.02822}{{\texttt{arXiv:1504.02822}}}.

\bibitem{Dittrich:2017hnl}
B.~Dittrich, C.~Goeller, E.~Livine, and A.~Riello, ``{Quasi-local holographic
  dualities in non-perturbative 3d quantum gravity I \textendash{} Convergence
  of multiple approaches and examples of Ponzano\textendash{}Regge statistical
  duals},'' Nucl. Phys. B {\bf 938} (2019) 807--877,
  \href{http://arXiv.org/abs/1710.04202}{{\texttt{arXiv:1710.04202}}}.

\bibitem{Turaev:1992hq}
V.~G. Turaev and O.~Y. Viro, ``{State sum invariants of 3 manifolds and quantum
  6j symbols},'' Topology {\bf 31} (1992) 865--902.

\bibitem{Witten:1988hc}
E.~Witten, ``{(2+1)-Dimensional Gravity as an Exactly Soluble System},'' Nucl.
  Phys. B {\bf 311} (1988) 46.

\bibitem{Witten:1988hf}
E.~Witten, ``{Quantum Field Theory and the Jones Polynomial},'' Commun. Math.
  Phys. {\bf 121} (1989) 351--399.

\bibitem{Dittrich:2016typ}
B.~Dittrich and M.~Geiller, ``{Quantum gravity kinematics from extended
  TQFTs},'' New J. Phys. {\bf 19} (2017), no.~1, 013003,
  \href{http://arXiv.org/abs/1604.05195}{{\texttt{arXiv:1604.05195}}}.

\bibitem{Dupuis:2017otn}
M.~Dupuis, L.~Freidel, and F.~Girelli, ``{Discretization of 3d gravity in
  different polarizations},'' Phys. Rev. D {\bf 96} (2017), no.~8, 086017,
  \href{http://arXiv.org/abs/1701.02439}{{\texttt{arXiv:1701.02439}}}.

\bibitem{Livine:2016vhl}
E.~R. Livine, ``{3d Quantum Gravity: Coarse-Graining and $q$-Deformation},''
  Annales Henri Poincare {\bf 18} (2017), no.~4, 1465--1491,
  \href{http://arXiv.org/abs/1610.02716}{{\texttt{arXiv:1610.02716}}}.

\bibitem{Reshetikhin:1990pr}
N.~Y. Reshetikhin and V.~G. Turaev, ``{Ribbon graphs and their invariants
  derived from quantum groups},'' Commun. Math. Phys. {\bf 127} (1990) 1--26.

\bibitem{Capovilla:2001zi}
R.~Capovilla, M.~Montesinos, V.~A. Prieto, and E.~Rojas, ``{BF gravity and the
  Immirzi parameter},'' Class. Quant. Grav. {\bf 18} (2001) L49--L52,
  \href{http://arXiv.org/abs/gr-qc/0102073}{{\texttt{arXiv:gr-qc/0102073}}}.
  [Erratum: Class.Quant.Grav. 18, 1157 (2001)].

\bibitem{Livine:2001jt}
R.~E. Livine and D.~Oriti, ``{Barrett-Crane spin foam model from generalized BF
  type action for gravity},'' Phys. Rev. D {\bf 65} (2002) 044025,
  \href{http://arXiv.org/abs/gr-qc/0104043}{{\texttt{arXiv:gr-qc/0104043}}}.

\bibitem{DePietri:1998hnx}
R.~De~Pietri and L.~Freidel, ``{so(4) Plebanski action and relativistic spin
  foam model},'' Class. Quant. Grav. {\bf 16} (1999) 2187--2196,
  \href{http://arXiv.org/abs/gr-qc/9804071}{{\texttt{arXiv:gr-qc/9804071}}}.

\bibitem{Holst:1995pc}
S.~Holst, ``{Barbero's Hamiltonian derived from a generalized Hilbert-Palatini
  action},'' Phys. Rev. D {\bf 53} (1996) 5966--5969,
  \href{http://arXiv.org/abs/gr-qc/9511026}{{\texttt{arXiv:gr-qc/9511026}}}.

\bibitem{Perez:2005pm}
A.~Perez and C.~Rovelli, ``{Physical effects of the Immirzi parameter},'' Phys.
  Rev. D {\bf 73} (2006) 044013,
  \href{http://arXiv.org/abs/gr-qc/0505081}{{\texttt{arXiv:gr-qc/0505081}}}.

\bibitem{Freidel:2005sn}
L.~Freidel, D.~Minic, and T.~Takeuchi, ``{Quantum gravity, torsion, parity
  violation and all that},'' Phys. Rev. D {\bf 72} (2005) 104002,
  \href{http://arXiv.org/abs/hep-th/0507253}{{\texttt{arXiv:hep-th/0507253}}}.

\bibitem{Freidel:2012np}
L.~Freidel and S.~Speziale, ``{On the relations between gravity and BF
  theories},'' SIGMA {\bf 8} (2012) 032,
  \href{http://arXiv.org/abs/1201.4247}{{\texttt{arXiv:1201.4247}}}.

\bibitem{Krasnov:2009iy}
K.~Krasnov, ``{Gravity as BF theory plus potential},'' Int. J. Mod. Phys. A
  {\bf 24} (2009) 2776--2782,
  \href{http://arXiv.org/abs/0907.4064}{{\texttt{arXiv:0907.4064}}}.

\bibitem{Freidel:2005ak}
L.~Freidel and A.~Starodubtsev, ``{Quantum gravity in terms of topological
  observables},''
  \href{http://arXiv.org/abs/hep-th/0501191}{{\texttt{arXiv:hep-th/0501191}}}.

\bibitem{Crane:1993if}
L.~Crane and D.~Yetter, ``{A Categorical construction of 4-D topological
  quantum field theories},''
\newblock 3, 1993.
\newblock
  \href{http://arXiv.org/abs/hep-th/9301062}{{\texttt{arXiv:hep-th/9301062}}}.

\bibitem{Pfeiffer:2001yd}
H.~Pfeiffer, ``{Four-dimensional lattice gauge theory with ribbon categories
  and the Crane-Yetter state sum},'' J. Math. Phys. {\bf 42} (2001) 5272--5305,
  \href{http://arXiv.org/abs/hep-th/0106029}{{\texttt{arXiv:hep-th/0106029}}}.

\bibitem{Alexandrov:2002br}
S.~Alexandrov and E.~R. Livine, ``{SU(2) loop quantum gravity seen from
  covariant theory},'' Phys. Rev. D {\bf 67} (2003) 044009,
  \href{http://arXiv.org/abs/gr-qc/0209105}{{\texttt{arXiv:gr-qc/0209105}}}.

\bibitem{Barrett:1999qw}
J.~W. Barrett and L.~Crane, ``{A Lorentzian signature model for quantum general
  relativity},'' Class. Quant. Grav. {\bf 17} (2000) 3101--3118,
  \href{http://arXiv.org/abs/gr-qc/9904025}{{\texttt{arXiv:gr-qc/9904025}}}.

\bibitem{Perez:2000ec}
A.~Perez and C.~Rovelli, ``{Spin foam model for Lorentzian general
  relativity},'' Phys. Rev. D {\bf 63} (2001) 041501,
  \href{http://arXiv.org/abs/gr-qc/0009021}{{\texttt{arXiv:gr-qc/0009021}}}.

\bibitem{Christensen:2009bi}
J.~D. Christensen, I.~Khavkine, E.~R. Livine, and S.~Speziale, ``{Sub-leading
  asymptotic behaviour of area correlations in the Barrett-Crane model},''
  Class. Quant. Grav. {\bf 27} (2010) 035012,
  \href{http://arXiv.org/abs/0908.4476}{{\texttt{arXiv:0908.4476}}}.

\bibitem{Barrett:2009mw}
J.~W. Barrett, R.~J. Dowdall, W.~J. Fairbairn, F.~Hellmann, and R.~Pereira,
  ``{Lorentzian spin foam amplitudes: Graphical calculus and asymptotics},''
  Class. Quant. Grav. {\bf 27} (2010) 165009,
  \href{http://arXiv.org/abs/0907.2440}{{\texttt{arXiv:0907.2440}}}.

\bibitem{Christodoulou:2023psv}
M.~Christodoulou, F.~D'Ambrosio, and C.~Theofilis, ``{Geometry Transition in
  Spinfoams},''
\newblock 2, 2023.
\newblock \href{http://arXiv.org/abs/2302.12622}{{\texttt{arXiv:2302.12622}}}.

\bibitem{Han:2024ydv}
M.~Han, H.~Liu, D.~Qu, F.~Vidotto, and C.~Zhang, ``{Cosmological Dynamics from
  Covariant Loop Quantum Gravity with Scalar Matter},''
  \href{http://arXiv.org/abs/2402.07984}{{\texttt{arXiv:2402.07984}}}.

\bibitem{Banburski:2014cwa}
A.~Banburski, L.-Q. Chen, L.~Freidel, and J.~Hnybida, ``{Pachner moves in a 4d
  Riemannian holomorphic Spin Foam model},'' Phys. Rev. D {\bf 92} (2015),
  no.~12, 124014,
  \href{http://arXiv.org/abs/1412.8247}{{\texttt{arXiv:1412.8247}}}.

\bibitem{Rovelli:2005yj}
C.~Rovelli, ``{Graviton propagator from background-independent quantum
  gravity},'' Phys. Rev. Lett. {\bf 97} (2006) 151301,
  \href{http://arXiv.org/abs/gr-qc/0508124}{{\texttt{arXiv:gr-qc/0508124}}}.

\bibitem{Livine:2006ab}
E.~R. Livine, S.~Speziale, and J.~L. Willis, ``{Towards the graviton from
  spinfoams: Higher order corrections in the 3-D toy model},'' Phys. Rev. D
  {\bf 75} (2007) 024038,
  \href{http://arXiv.org/abs/gr-qc/0605123}{{\texttt{arXiv:gr-qc/0605123}}}.

\bibitem{Alesci:2008ff}
E.~Alesci, E.~Bianchi, and C.~Rovelli, ``{LQG propagator: III. The New
  vertex},'' Class. Quant. Grav. {\bf 26} (2009) 215001,
  \href{http://arXiv.org/abs/0812.5018}{{\texttt{arXiv:0812.5018}}}.

\bibitem{Rovelli:2008aa}
C.~Rovelli and F.~Vidotto, ``{Stepping out of Homogeneity in Loop Quantum
  Cosmology},'' Class. Quant. Grav. {\bf 25} (2008) 225024,
  \href{http://arXiv.org/abs/0805.4585}{{\texttt{arXiv:0805.4585}}}.

\bibitem{Livine:2011up}
E.~R. Livine and M.~Martin-Benito, ``{Classical Setting and Effective Dynamics
  for Spinfoam Cosmology},'' Class. Quant. Grav. {\bf 30} (2013) 035006,
  \href{http://arXiv.org/abs/1111.2867}{{\texttt{arXiv:1111.2867}}}.

\bibitem{Gozzini:2019nbo}
F.~Gozzini and F.~Vidotto, ``{Primordial Fluctuations From Quantum Gravity},''
  Front. Astron. Astrophys. Cosmol. {\bf 7} (2021) 629466,
  \href{http://arXiv.org/abs/1906.02211}{{\texttt{arXiv:1906.02211}}}.

\bibitem{Dittrich:2013xwa}
B.~Dittrich and S.~Steinhaus, ``{Time evolution as refining, coarse graining
  and entangling},'' New J. Phys. {\bf 16} (2014) 123041,
  \href{http://arXiv.org/abs/1311.7565}{{\texttt{arXiv:1311.7565}}}.

\bibitem{Bahr:2016hwc}
B.~Bahr and S.~Steinhaus, ``{Numerical evidence for a phase transition in 4d
  spin foam quantum gravity},'' Phys. Rev. Lett. {\bf 117} (2016), no.~14,
  141302, \href{http://arXiv.org/abs/1605.07649}{{\texttt{arXiv:1605.07649}}}.

\bibitem{Steinhaus:2020lgb}
S.~Steinhaus, ``{Coarse Graining Spin Foam Quantum Gravity\textemdash{}A
  Review},'' Front. in Phys. {\bf 8} (2020) 295,
  \href{http://arXiv.org/abs/2007.01315}{{\texttt{arXiv:2007.01315}}}.

\bibitem{Asante:2019lki}
S.~K. Asante, B.~Dittrich, F.~Girelli, A.~Riello, and P.~Tsimiklis, ``{Quantum
  geometry from higher gauge theory},'' Class. Quant. Grav. {\bf 37} (2020),
  no.~20, 205001,
  \href{http://arXiv.org/abs/1908.05970}{{\texttt{arXiv:1908.05970}}}.

\bibitem{Rivasseau:2011hm}
V.~Rivasseau, ``{Quantum Gravity and Renormalization: The Tensor Track},'' AIP
  Conf. Proc. {\bf 1444} (2012), no.~1, 18--29,
  \href{http://arXiv.org/abs/1112.5104}{{\texttt{arXiv:1112.5104}}}.

\bibitem{Gurau:2012vu}
R.~Gurau, ``{A review of the large N limit of tensor models},''
  \href{http://arXiv.org/abs/1209.4295}{{\texttt{arXiv:1209.4295}}}.

\bibitem{Oriti:2016qtz}
D.~Oriti, L.~Sindoni, and E.~Wilson-Ewing, ``{Emergent Friedmann dynamics with
  a quantum bounce from quantum gravity condensates},'' Class. Quant. Grav.
  {\bf 33} (2016), no.~22, 224001,
  \href{http://arXiv.org/abs/1602.05881}{{\texttt{arXiv:1602.05881}}}.

\bibitem{Marchetti:2022nrf}
L.~Marchetti, D.~Oriti, A.~G.~A. Pithis, and J.~Th\"urigen, ``{Mean-Field Phase
  Transitions in Tensorial Group Field Theory Quantum Gravity},'' Phys. Rev.
  Lett. {\bf 130} (2023), no.~14, 141501,
  \href{http://arXiv.org/abs/2211.12768}{{\texttt{arXiv:2211.12768}}}.

\bibitem{Asante:2021zzh}
S.~K. Asante, B.~Dittrich, and J.~Padua-Arguelles, ``{Effective spin foam
  models for Lorentzian quantum gravity},'' Class. Quant. Grav. {\bf 38}
  (2021), no.~19, 195002,
  \href{http://arXiv.org/abs/2104.00485}{{\texttt{arXiv:2104.00485}}}.

\bibitem{Asante:2022dnj}
S.~K. Asante, B.~Dittrich, and S.~Steinhaus, ``{Spin foams, Refinement limit
  and Renormalization},''
  \href{http://arXiv.org/abs/2211.09578}{{\texttt{arXiv:2211.09578}}}.

\bibitem{Dittrich:2022yoo}
B.~Dittrich and A.~Kogios, ``{From spin foams to area metric dynamics to
  gravitons},'' Class. Quant. Grav. {\bf 40} (2023), no.~9, 095011,
  \href{http://arXiv.org/abs/2203.02409}{{\texttt{arXiv:2203.02409}}}.

\bibitem{Han:2023cen}
M.~Han, H.~Liu, and D.~Qu, ``{Complex critical points in Lorentzian spinfoam
  quantum gravity: Four-simplex amplitude and effective dynamics on a
  double-\ensuremath{\Delta}3 complex},'' Phys. Rev. D {\bf 108} (2023), no.~2,
  026010, \href{http://arXiv.org/abs/2301.02930}{{\texttt{arXiv:2301.02930}}}.

\bibitem{Gozzini:2021mex}
F.~Gozzini, {\em {Spin foam models of quantum gravity : advances through new
  techniques and numerical codes}}.
\newblock PhD thesis, Aix-Marseille U., 2021.

\bibitem{Dona:2022dxs}
P.~Dona and P.~Frisoni, ``{How-to Compute EPRL Spin Foam Amplitudes},''
  Universe {\bf 8} (2022), no.~4, 208,
  \href{http://arXiv.org/abs/2202.04360}{{\texttt{arXiv:2202.04360}}}.

\bibitem{Dona:2022yyn}
P.~Dona, M.~Han, and H.~Liu, {\em Spinfoams and High-Performance Computing},
  pp.~1--38.
\newblock Springer Nature Singapore, Singapore, 2023.
\newblock \href{http://arXiv.org/abs/2212.14396}{{\texttt{arXiv:2212.14396}}}.

\end{thebibliography}\endgroup

\end{document}